\documentclass{amsart}
\usepackage{amssymb}
\usepackage{amsfonts}
\usepackage{epsfig}
\usepackage{graphicx}
\newcounter{eqnletter}[equation]
\catcode`\@=11
\@addtoreset{equation}{section}
\catcode`\@=12

\newcommand{\Tr}[1]{{\sf tr}\left(#1\right)}

\begin{document} {\centerline {\LARGE {\bf Dilatation operator and Cayley
      graphs}}} \vskip 1 cm \centerline {M.~Bonini, G.M.~Cicuta and E.~Onofri }
\vskip .5 cm {\small \centerline {{\it Dipartimento di Fisica, Univ.  di Parma,
      v.~G.P.~Usberti 7A, 43100 Parma, Italy }} \centerline {{\it and INFN,
      Sezione di Milano Bicocca, Gruppo di Parma} \quad \quad } \centerline
  {\small{\sf email}:\emph{ name(at)fis.unipr.it}}}
 
\vskip .7 cm {\centerline{\bf Abstract}}
 We use the algebraic definition of the Dilatation operator provided by Minahan, Zarembo, Beisert, Kristijansen, Staudacher, proper for single trace products of scalar fields, at leading order in the large-$N$ 't Hooft limit
 to develop a new approach to the evaluation of the spectrum of the Dilatation operator. We discover a vast number of exact sequences of eigenstates.\\
  
\vskip .7 cm
\section{Introduction and Summary}
The past five years witnessed impressive amount of work and progress in the understanding of $\mathcal{ N}=4$ superconformal Yang Mills theory with gauge group $SU(N)$ in the large-$N$ limit.\\

\noindent
The most relevant results include the increasingly detailed correspondence
between states in this theory with string states on $AdS_5 \times S^5$, the
technical improvements in the evaluation of the anomalous dimension of operators
which led to the discovery of quantum integrability, some unexpected relations
with high energy sectors of quantum chromodynamics.  We cannot possibly quote
the pertinent vast literature and we refer the reader to the papers \cite{r1}
\cite{r2} \cite{ple} for introduction to the subject and to the original
literature.  Each of these outstanding results were obtained in the large-$N$ 't
Hooft limit, in several sectors of the theory, at several orders in loop
expansion, then suggesting the possibility of a complete understanding of the
theory.

It seems important both for a more complete understanding of the $\mathcal{
  N}=4$ Super Yang Mills $SU(N)$ theory at large-$N$ and for further tests of
the AdS/CFT Maldacena conjecture  to  evaluate
the eigenvalues of the dilatation operator for {\bf all} states product of a small number of fields 
 and for sequences made of an arbitrary number of fields.\\
To this goal, an essential progress was obtained by expressing the dilatation
operator in a way that translates the evaluation of anomalous conformal
dimension of states into a diagonalization problem in finite dimensional spaces,
then avoiding the previous cumbersome evaluation by Feynman graphs.\\

We shall only work in the sector of states given by products of the six real
scalar matrix fields $\phi_j$ of the theory . An operator containing the product
of $n$ matrix fields like $\Tr {\phi_1 \phi_2 \cdots \phi_n}$ has conformal
dimension $\bigtriangleup_0=n$, at tree level. We refer this number as the main
quantum number of the operator or state. In this work we limit ourselves to the
exact analytic evaluation of the one loop contribution $\bigtriangleup_2$, to
the conformal dimension for these single trace states in leading order in the
large-$N$ 't Hooft limit.
Another good quantum number is parity. One may define a parity
operator $P$ which inverts the order of the matrix fields inside a
single trace, $P\, \Tr {\phi_1 \phi_2 \cdots \phi_n}=\Tr {\phi_n
  \phi_{n-1} \cdots \phi_1}$. Since the Hamiltonian corresponding to
the Dilatation operator commutes with the parity operator, it is
possible and useful to have eigenvectors with definite parity.\\

One may consider a set of states obtained by permuting the positions of the complex fields $Z$, $W$, $Y$, later defined in  eq.(\ref{t.20}), inside a trace 
$\Tr {Z^aW^bY^c}$. The states of this set may be regarded as basis vectors of a linear vector space invariant under the action of the one-loop Dilatation operator. We refer to this vector space as the sector $\Tr {Z^a W^b Y^c}$.
The dimension of the vector space increases rapidly as the main quantum number $\bigtriangleup_0=a+b+c$ increases. For instance the sector $\Tr {Z^2W^2}$ has only $2$ independent states, both of positive parity, whereas the sector $\Tr {Z^3W^2Y^2}$ has $30$ independent states. To evaluate eigenvalues and eigenvectors of the Dilatation operator in a given sector it is very useful to  use basis vectors with definite parity obtained by sum and differences of pairs of the previous basis vectors, then splitting large matrices into two smaller ones.\\
This evaluations will be called direct diagonalization of the Dilatation operator. It is the simplest 
procedure provided the main quantum number $\bigtriangleup_0=a+b+c$ is a small integer.
We  provide a summary, in Appendix A, of all eigenstates of the dilatation
operator for low quantum number $4 \leq \bigtriangleup_0 \leq 7$. 
  This represents a useful
information for the understanding of the theory, for checking evaluations we
perform for arbitrary value of $\bigtriangleup_0$ and for any comparison with energy of states in
string theory.\\

Possibly the most important result of our paper is the evaluation of a vast
number of exact sequences of eigenstates of the dilatation operator, which
appear as eqs. ($C.1$) and ($C.2$). To our knowledge, these sequences
are not known in the literature. They have the same eigenvalues of the well
known sequences for two impurities, reproduced in eqs. ($D.1$) and
($D.2$), which may be said to belong to the sector $\Tr {Z^n \phi_a \phi_b}$
of the theory. Our sequences are valid for every sector of the form $\Tr
{Z^a W^{b} Y^{c}}$ with arbitrary values for the (non-negative) integer
exponents. For instance, in the sector $\Tr {Z^{n-2}W^2 Y}$, we provide very explicit expressions of exact sequences of eigenstates in eqs. ($D.3$), ($D.4$).\\

Our method is rather different from the powerful methods
(superconformal algebra, integrability and Bethe ansatz) used in
extensive evaluations already performed with the same goal. We define
an auxiliary Hamiltonian, which might be called pertinent to a
nearest-neighbor exchange model where the number of flavours of the
matrix fields $\phi_j$ is unlimited and in every configuration of the
fields the flavours are all distinct.
\noindent

Configurations are sums of permutations and the Hamiltonian changes permutations
into permutations. The Hamiltonian belongs to the group algebra and, not
surprisingly, the analysis of this auxiliary model leads the study of the
irreducible representations of the permutation group $S_n$, its Young
projectors, its group algebra. At the end we recover information pertinent
$\mathcal{ N}=4$ Super Yang Mills theory by trivial replacements in the results.\\

Our method may be used to discover new sequences of eigenstates of the dilatation operator, belonging to irreducible representations different from the one we studied. These would have different eigenvalues. In this case, we cannot
anticipate if the analytic evaluation can be carried to the end.\\

The outline of the paper is the following : in Section $2$ we define our auxiliary model and its analysis which leads us to the evaluation of eigenvalues and eigenvector for one element of the group algebra of the symmetric group $S_n$ in several irreducible representations. Every representation will be denoted by the sequence of integer numbers 
counting  the number of boxes in horizontal rows of the Young tableaux. The eigenvalues of the Hamiltonian in the auxiliary model include the eigenvalues of the Dilatation operator of the Super Yang
Mills theory. For a number of irreducible representations we collected them in
Appendix B.

The method to find the eigenvectors pertinent to specific representations is
described in Section 3 and in Appendix C where we obtain an explicit solution
for the representation $(n-1,1)$. This is the easiest non trivial representation
and contains the sequences we mentioned before, the well known ones and the new
ones.  Section 4 and Appendix D contain the easiest replacements for the
general sequences.

\section{Cayley graphs}
It was shown by Minahan and Zarembo \cite{mz} that the action of the dilatation operator on single-trace states of the  product of $n+1$ scalar matrix-fields, at one loop order in the large-$N$ 't Hooft limit, may be replaced by the matrix 
\begin{equation}
\Gamma=\frac{\lambda}{16 \pi^2} \sum_{l=1}^{n+1} \left(K_{l,l+1}+2-2P_{l,l+1}\right) \quad , \quad \lambda=g^2_{YM}N
 \label{t.1}
\end{equation}
The two operators $K_{l,l+1}$, $P_{l,l+1}$ act only on the pair of fields in the positions $(l,l+1)$ inside the single-trace string of $n+1$ scalar fields.\\
The operator $K_{l,l+1}$, called a trace operator, is
$$K_{l,l+1} \, \Tr {\phi_{\alpha_1}\cdots \phi_{\alpha_{l-1}}\phi_{\alpha_l} \phi_{\alpha_{l+1}}\phi_{\alpha_{l+2}}
\cdots \phi_{\alpha_{n+1}} }=
\delta_{\alpha_l\,,\,\alpha_{l+1} }\sum_{k=1}^6
\Tr {\phi_{\alpha_1}\cdots \phi_{\alpha_{l-1}} \phi_k\phi_k  \phi_{\alpha_{l+2}}\cdots \phi_{\alpha_{n+1}} }$$
It yields zero if the pair of matrix fields at positions $l$ and $l+1$ have different flavour.\\
The operator $P_{l,l+1}$ exchanges the flavour of the matrix fields at positions $l$ and $l+1$ irrespective of the flavours being equal or different
$$P_{l,l+1} \, \Tr {\phi_{\alpha_1}\cdots \phi_{\alpha_{l-1}}\phi_{\alpha_l} \phi_{\alpha_{l+1}}\phi_{\alpha_{l+2}}
\cdots \phi_{\alpha_{n+1}} }=
\Tr {\phi_{\alpha_1}\cdots \phi_{\alpha_{l-1}}\phi_{\alpha_{l+1}} \phi_{\alpha_l}\phi_{\alpha_{l+2}}
\cdots \phi_{\alpha_{n+1}} }$$
Of course the trace operator may be neglected if the dilatation operator acts on configurations $  \Tr {\phi_{\alpha_1} \phi_{\alpha_2}\cdots \phi_{\alpha_{n+1}} }$ where the flavours $\alpha_j$ of the matrix fields are all different. This is possible only for short chains $(n+1 \leq 6)$ and leads us to the definition of a auxiliary model.\\

Our auxiliary model is defined by the generalized dilatation operator
\begin{equation}
 \Gamma=\frac{\lambda}{8 \pi^2} \,{\bf L} \quad , \quad
{\bf L}= \sum_{l=1}^{n+1} \left(I_{l,l+1}-P_{l,l+1}\right) ={\bf I}-{\bf A} \qquad ,
   \qquad \lambda=g^2_{YM}N
 \label{t.2}
\end{equation}
acting on operators $\Tr {\phi_{\alpha_1} \phi_{\alpha_2}\cdots
  \phi_{\alpha_{n+1}} }$ where the flavours $\alpha_j$ of the matrix fields are
all different, that is $\{\alpha_1 , \alpha_2 ,..,\alpha_{n+1}\}$ is a
permutation of the set of integers $\{1,2,..,n+1\}$. Because of the cyclic
property of trace, the number of states is $n!$ . One may fix the
position of one flavour, let us choose the first, and consider the set of
independent states $\Tr{ \phi_1\phi_{\alpha_2}..\phi_{\alpha_{n+1}} }$ where the
sequence $\{\alpha_2 , \alpha_3 ,..,\alpha_{n+1}\}$ is a permutation of the
permutation group $S_n$ acting on the sequence $\{2,3,..,n+1\}$ .\\

Next we proceed to evaluate eigenvalues and eigenstates for the auxiliary model. It might seem that such spectrum would provide the correct spectrum of the dilatation operator in superconformal Yang Mills theory only for short chains and just in the sector where all flavours are different. We suggest in
the last section that a simple replacement rule allows us to recover from the
analysis of the auxiliary model the corresponding information for super Yang
Mills in the sector $\Tr {Z^a W^b Y^c}$ for any choice of integers $a,b,c$.
   
One may consider the $n!$ states 
 $\Tr { \phi_1\phi_{\alpha_2}..\phi_{\alpha_{n+1}} } $ as a basis in a  vector space $V_{n!}$. 
 With the above convention of fixing flavour one in first place, 
the form of the operator ${\bf A}= \sum_{l=1}^{n+1} P_{l,l+1}$ , writing the permutations as cycles, is
\begin{equation}
{\bf A}= \sum_{l=1}^{n+1} P_{l,l+1}=(2,3,..,n,n+1)+(2,3)+..(n,n+1)+(n+1,n,..,3,2)
 \label{t.3}
\end{equation}
The first permutation is the inverse of the last one, whereas each transposition
coincides with its inverse.  \footnote{ {\bf Warning about conventions.} Some
  care is necessary when we translate results from the theory of representations
  of the symmetric group to the present generalized Heisenberg model on a chain.\\
  We have defined a one-to-one correspondence between states with
  $\bigtriangleup_0=n+1$ and elements of the permutation group $S_n$ :
\begin{equation}
 {\rm tr}\,[\phi_1 \phi_{\alpha_2} \cdots\phi_{\alpha_{n+1}}] \sim \left( 2 \, 3 \, \cdots n+1 \atop \alpha_2 \, \alpha_3 ..\alpha_{n+1}\right) \qquad \label{f.1}
 \end{equation}
However products of exchange operators $P_{l,l+1}$ act in reverse order of the usual conventions on products of permutations. An example will illustrate it :
\begin{eqnarray}
P_{4,5}P_{3,4} {\rm tr}\,[\phi_1 \phi_5 \phi_4 \phi_3 \phi_2]=P_{4,5}  {\rm tr}\,[\phi_1 \phi_5 \phi_3 \phi_4 \phi_2]=
 {\rm tr}\,[\phi_1 \phi_5 \phi_3 \phi_2 \phi_4]\nonumber
 \end{eqnarray}
 It corresponds to :
 \begin{eqnarray}
\left( 1 \, 2 \, 3 \, 4 \, 5 \atop 1 \, 5 \, 4 \, 3 \, 2 \right) (34)(45)= \left( 1 \, 2 \, 3 \, 4 \, 5 \atop 1 \, 5 \, 3 \, 2 \, 4 \right)  \nonumber
 \end{eqnarray}
 As a general rule, we use the theory of irreducible representations of $S_n$,
 the projectors related to Young tableaux, as in Appendix B and C, with the
 generally used conventions on products of group elements. Often we consider
 linear combinations of group elements, that is elements in the group ring. The
 results may be translated into linear combinations of traces of products of
 matrix fields by first taking the inverse of each permutation, then applying
 the correspondence (\ref{f.1}). }\\

The operator ${\bf A}$ is represented as a real symmetric matrix in the space
$V_{n!}$ and it is the adjacency matrix of the graph $G(V,E)$ associated to the
matrix. The set of vertices is the set of the $n!$ independent states $\Tr{
  \phi_1\phi_{\alpha_2}..\phi_{\alpha_{n+1} } }$ , a link connects vertex $v_i$
with vertex $v_k$ if one of the $n+1$ permutations $\pi$ in the sum in
eq.(\ref{t.3}), is such that $v_k=\pi v_i$.\\

The operator ${\bf L}$  given in eq.(\ref{t.2}) 
  is the Laplacian of the graph.  Since every vertex in the graph has
the same degree $n+1$, the spectrum of the adjacency matrix ${\bf A}$ is
trivially related to the spectrum of the Laplacian matrix ${\bf L}$. The
eigenvalues of ${\bf L}$, here called \boldmath $\bigtriangleup_2, $\unboldmath
\, provide the one loop contribution to the anomalous dimension
  \boldmath $$
\bigtriangleup=\bigtriangleup_0+\mbox{\unboldmath
  $\frac{\lambda}{8\pi^2}$}\bigtriangleup_2 \quad , \quad
\bigtriangleup_0=n+1 \,.$$ \unboldmath

Let us recall \cite{big} that a graph $G(V,E)$ where the set of vertices is the set of elements of a group, and the set of edges is a subset of the previous set, provided it is closed under taking the inverse, is a Cayley graph. Then the graph we are discussing in this section is the Cayley graph on the group of permutations $S_n$ with the set of $n+1$ connections listed on the right side of eq.(\ref{t.3}). The evaluation of the spectrum of the adjacency matrix of Cayley graphs even for large graphs, is greatly facilitated by its symmetries.\\

We recall that for any $n$ there exist a very easy representation of the permutation group $S_n$ of degree $n!$ . It is obtained by considering the elements $g \in S_n$ both as basis vectors  as well as operators in the vector space spanned by the basis vectors. Each 
$g \in S_n$ is represented by a matrix with only one entry equal to one and the remaining entries equal to zero in each row and in each column. This representation is sometimes called the {\bf regular} representation.\\
The real symmetric matrix ${\bf A}$ , of order $n!$ in eq.(\ref{t.3}) is the sum of $(n+1)$ real symmetric matrices which are the regular representation of $n-1$ transpositions and $2$ long cycles in $S_n$. It contains
$n+1$ entries equal to one in each row and in each column. \\
Any set of matrices which are the regular representation of a set of elements $g \in S_n$, allow a simultaneous block decomposition, where the matrices in the blocks are the irreducible representations of the elements $g \in S_n$.
A irreducible representation of degree $f$ occurs $f$ times in this decomposition \cite{boe}, then the spectrum of ${\bf A}$ is given by the eigenvalues of ${\bf A}$ in the irreducible representation of degree $f$ with multiplicity $f$ (times the multiplicity of the eigenvalue in the irreducible).\\
Since the regular representation is a matrix of order $n!$, the size increases too rapidly to allow a direct evaluation of eigenvalues and eigenvectors of the matrix {\bf A} beyond the smallest values of $n$. But every information on the spectrum is recovered by the analysis of the irreducible representations occurring in the block decomposition of the regular representation.\\

The evaluation of the eigenvalues does not need writing the irreducible representation for the $n-1$ transpositions and the $2$ cycles occurring in eq.(\ref{t.3}). Indeed from the knowledge of the characters of all classes of elements of $S_n$ in a given representation of degree $f$, one obtains the character of the matrices $A$, $A^2$, \dots , $A^f$ then the characteristic equation for the matrix $A$  then its eigenvalues. However we found easier to profit from the explicit
 irreducible representations tabulated for the generators of the permutation group $S_n$  explicitly listed \cite{enc} up to some value of $n$. \\

\boldmath $ \bigtriangleup_0=4.$\unboldmath \, By this method we  evaluate the $6$ eigenvalues of ${\bf A}$ for $n=3$ : \{
$\lambda=4$ singlet, $\lambda=0$ with multiplicity $3$, $\lambda=-2$ with multiplicity $2$ \}. They translate respectively to the eigenvalues of the Laplacian operator : \{$\mathbf{\bigtriangleup_2=0}$ singlet, 
\boldmath $\bigtriangleup_2=4$ \unboldmath \,with multiplicity $3$,  \boldmath $\bigtriangleup_2=6$ \unboldmath \, with multiplicity $2$ \}. The representation of dimension $6$ is partitioned into the irreducible representations $6=1+1+2^2$. The first singlet, corresponding to the identity representation is the totally symmetric eigenstate, the second singlet corresponds to the alternate representation, finally the $2$-dimensional irreducible representation provides $\lambda=0$ and $\lambda=-2$.\\
In a more symmetric fashion, the $3$ eigenstates with $\lambda=0$ may be chosen as
\begin{eqnarray}
u_1 &=& \Tr{\phi_1\phi_2\phi_3\phi_4-\phi_1\phi_4\phi_3\phi_2} \qquad, \nonumber \\
u_2 &=& \Tr{\phi_1\phi_2\phi_4\phi_3-\phi_1\phi_3\phi_4\phi_2} \qquad, \nonumber \\
u_3 &=& \Tr{\phi_1\phi_4\phi_2\phi_3-\phi_1\phi_3\phi_2\phi_4} \qquad 
  \label{t.3a}
\end{eqnarray}
The $2$ eigenstates with $\lambda=-2$ may be chosen as
\begin{eqnarray}
u_4 &=& \Tr{ \phi_1\phi_2\phi_3\phi_4-\phi_1\phi_2\phi_4\phi_3+
\phi_1\phi_4\phi_3\phi_2-\phi_1\phi_3\phi_4\phi_2} 
\qquad, \nonumber \\
u_5 &=& \Tr{ \phi_1\phi_2\phi_4\phi_3-\phi_1\phi_3\phi_2\phi_4+
\phi_1\phi_3\phi_4\phi_2-\phi_1\phi_4\phi_2\phi_3} 
  \label{t.3b}
\end{eqnarray}
\vskip 10 pt
\boldmath $\bigtriangleup_0=5.$ \unboldmath \, The permutation group $S_4$ has $5$ irreducible representations : two of degree $1$, 
one $2$-dimensional and two of degree $3$. Correspondingly the matrix ${\bf A}$, of order $24$ decomposes into blocks, $24=1+1+2^2+2 \times 3^2$.\\ The eigenvalues of the two $1$-dimensional representations are $\lambda=\pm 5$.
The eigenvalues associated to the other three irreducible representations are exhibited in the Appendix B. Eigenvalues and multiplicities are collected in Table $1$.\\

\begin{center}
  \begin{tabular}{|c||c| c|c|c|c|c||} \hline
   $\lambda $ & $5$ & $ \sqrt{5} $ & $1$ & $-1$ &  $ -\sqrt{5} $ & $-5$ \\ \hline
$\bigtriangleup_2$  & 0 & $5-\sqrt{5}$ & 4 & 6 & $5+\sqrt{5}$ & $10 $\\  \hline
multiplicity           & 1 & 6   & 5 & 5 & 6 & 1\\ \hline
\end{tabular} \\  

 \small{Table $1$. Eigenvalues of the states $\Tr {\phi_1 \phi_{\alpha_2}\phi_{\alpha_3}\phi_{\alpha_4} \phi_{\alpha_5} }$ ,
  $\bigtriangleup_0=5$ has multiplicity$=24$. The spectrum of the Laplacian has the symmetry $\bigtriangleup_2\to 10-  \bigtriangleup_2$ } \\
\end{center}
\vskip 8 pt
   The spectrum of $\bf A$ is symmetric with respect to the origin because all the $5$ permutations occurring in eq.(\ref{t.3}) may be decomposed in a odd number of transpositions, then the graph with $24$ vertices each of degree $5$ is bipartite.\\ 
We omit listing the eigenvectors in the degenerate subspaces as they may be chosen in several ways. \\

\boldmath $\bigtriangleup_0=6.$ \unboldmath \, The permutation group $S_5$ has $7$ irreducible representations : two of degree $1$, 
two of degree $4$,  two of degree $5$, one $6$-dimensional. Correspondingly the matrix ${\bf A}$, of order $120$ decomposes into blocks $120=1+1+2\times 4^2+2 \times 5^2+6^2$.\\
The eigenvalues of the two $1$-dimensional representations are $\lambda=6$, $\lambda=-2$. The eigenvalues associated to the other five irreducible representations are exhibited in the Appendix B. Eigenvalues and multiplicities are collected in Table $2$.
\vskip 5 pt
\begin{center}
 \begin{tabular}{|c||c| c|c|c|c|c|c|c|c|c||} \hline
  $\lambda $ & $6$ & $4$ & $1+\sqrt{5} $ & $-1+\sqrt{13}$ & $1$ & $0$ &  $ 1-\sqrt{5} $ & $-2$ & $-4$ &$-1-\sqrt{13}$ \\ \hline
$\bigtriangleup_2$  & 0 & 2 & $5-\sqrt{5}$ & $7-\sqrt{13}$ & 5 & 6 & $5+\sqrt{5}$ & $8 $ & 10 & $7+\sqrt{13}$\\  \hline
multiplicity           & 1&10& 9 & 5     & 32 & 15 & 9 & 25 & 9 & 5\\ \hline
\end{tabular} \\
 \small{Table $2$. Eigenvalues of the states
 $\Tr {\phi_1  \phi_{\alpha_2}\phi_{\alpha_3}\phi_{\alpha_4}\phi_{\alpha_5} \phi_{\alpha_6} }$ , \\
 $\bigtriangleup_0=6$ has multiplicity$=120$.}\\
 \end{center}
 \vskip 10 pt
\boldmath $\bigtriangleup_0=7.$ \unboldmath \, The permutation group $S_6$ has $11$ irreducible representations : two of degree $1$, 
four of degree $5$,  two of degree $9$, two of degree $10$, one $16$-dimensional. Correspondingly the matrix ${\bf A}$, of order $720$ decomposes into blocks $720=1+1+4\times 5^2+2 \times 9^2+2 \times 10^2+16^2$.\\
The eigenvalues of the two $1$-dimensional representations are $\lambda=\pm 7$. The eigenvalues associated to the other nine irreducible representations are exhibited in the Appendix B. Eigenvalues and multiplicities are collected in Table $3$.\\

 \begin{center}
 \begin{tabular}{|c||c| c|c|c|c|c|c|c|c|c|} \hline
 $\lambda$ &  $7$ & $7-8\sin^2(\pi/7)$ & $5$ & $7-y_1$ & $7-z_1$ & $3$ & $7-8\sin^2(2\pi/7)$ & $2$  & $7-z_2$ & $1$  \\  \hline
$\bigtriangleup_2$  & $0$ & $8\sin^2(\pi/7)$ & $2$ & $y_1$ & $z_1$ & $4$ & $8\sin^2(2\pi/7)$ & 5  & $z_2$ & $6$  \\  \hline
multiplicity           & 1&15& 14 & 21  & 21 & 70 & 15 & 28 & 21 & 119 \\ \hline
\end{tabular} 
\vskip 1 cm

 \begin{tabular}{|c||c| c|c|c|c|c|c|} \hline
 $\lambda$ & $-7+8 \sin^2(3\pi/7)$ & $0$ & $7-8\sin^2(3\pi/7)$  & $-1$ & $7-y_2$ & $-2$ & $-7+8\sin^2(2\pi/7)$  \\  \hline
$\bigtriangleup_2$  & $14-8 \sin^2(3\pi/7)$ & 7 & $8\sin^2(3\pi/7)$  & $8$ & $y_2$ & 9 & $14-8\sin^2(2\pi/7)$  \\  \hline
multiplicity           & 15 & 40 & 15 & 119  & 21 & 28 & 15 \\ \hline
\end{tabular} 
\vskip 1 cm

\begin{tabular}{|c||c| c|c|c|c|c||} \hline
$\lambda$ & $-3$ & $7-y_3$ & $7-z_3$ & $-5$ & $-7+8\sin^2(\pi/7)$ & $-7$   \\  \hline
$\bigtriangleup_2$  & 10 & $y_3$ & $z_3$ & 12 & $14-8\sin^2(\pi/7)$ & 14   \\  \hline
multiplicity          & 70 & 21 & 21  & 14 & 15 &  1 \\ \hline
\end{tabular} 
\vskip .5 cm
\small{Table $3$. Eigenvalues of the states $\Tr {\phi_{\alpha_1} \phi_{\alpha_2}\phi_{\alpha_3}\phi_{\alpha_4}\phi_{\alpha_5} \phi_{\alpha_6} \phi_{\alpha_7}}$  , \\
 $\bigtriangleup_0=7$ has multiplicity $=720$.\\
$z_j$ are the roots of the cubic $E^3-20 E^2+116 E-200=0$\\
$y_j$ are the roots of the cubic $E^3-22 E^2+144 E-248=0$\\
   The spectrum  is symmetric with respect $\bigtriangleup_2=7$.  }
\end{center}
  \vskip 1 cm

Appendix B collects all the irreducible representations for {\bf A} and its eigenvalues up to $\bigtriangleup_0=7$.

\section{The sequences}
 
For any irreducible representation, of degree $f$, one may find a suitable set of $f$ basis vectors and represent the operator {\bf A} in eq.(\ref{t.3}) as matrix of order $f$, then finding the eigenstates of the dilatation operator.
In this section we show the method for a simple example of fixed order. In Appendix C the method is generalized to the much more interesting case of the easiest representations for $S_n$ then obtaining sequences of eigenvectors for any $n$.
One finds in the literature \cite{boe} the method to evaluate the basis vectors useful for any irreducible representation: they are the left cosets  obtained by multiplying each $g \in S_n$ times the Young projector $Y=PQ$ associated to the given irreducible representation.\\

We illustrate the method with an example. Let us consider the permutation group $S_4$ and the irreducible representation associated to the partition $(3,1)$ which has degree $f=3$. The Young  tableau in the figure \\
\epsfig{file=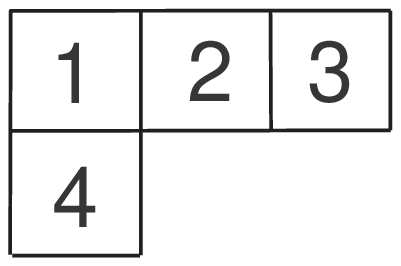 ,height=2cm}\\
is associated to the Young projector $Y=PQ$, where $P$ is the totally symmetric projector $P=e+(12)+(13)+(23)+(123)+(132)$
and $Q$ is the antisymmetric projector $Q=e-(14)$.\\
Three basis vectors are determined by the products $g \cdot Y$ , with $g \in S_4$. The evaluation is shortened by a couple of simple remarks. Each $g \in S_4$ is written in one of the four sets $a_j$ , $j=1,2,3,4$
 \begin{eqnarray}
a_1&=& (14) \cdot \left\{ e,(12), (13), (23), (123), (132) \right\} \quad , \nonumber\\
a_2&=& (24) \cdot \left\{ e,(12), (13), (23), (123), (132) \right\} \quad , \nonumber\\
a_3&=& (34) \cdot \left\{ e,(12), (13), (23), (123), (132) \right\} \quad , \nonumber\\
a_4&=& e \cdot \left\{ e,(12), (13), (23), (123), (132) \right\} \quad  \nonumber
\end{eqnarray} 
Furthermore if $g \in S_3 \in S_4$, $g \cdot PQ=PQ$. Then the $24$ products $g \cdot PQ$ for $g \in S_4$ fall into four cosets
\begin{eqnarray}
v_1&=& (14) \cdot  PQ \quad , \nonumber\\
v_2&=& (24) \cdot  PQ \quad , \nonumber\\
v_3&=& (34) \cdot  PQ \quad , \nonumber\\
v_4&=&  PQ \quad  \label{t.8}
\end{eqnarray}
It is also clear that $v_1+v_2+v_3+v_4=0$ because that sum is the product of the totally symmetric projector in $S_4$ times $Q$. Then a basis of $3$ independent (although not orthogonal) vectors for the irreducible representation corresponding to the partition $(3,1)$ may be chosen as three among the four $v_j$. Let us express $v_3=-(v_1+v_2+v_4)$.\\
The action of the operator $
A=(12)+(23)+(34)+(1234)+(4321)$ on the three basis vectors $v_j$ is

 \begin{eqnarray}
 A \, v_1 &=&  2v_1+2v_2+v_4 \qquad \nonumber\\
A\,v_2 &=&2 v_1+v_2 +2v_3=-v_2-2v_4 \qquad \nonumber\\
A \,v_4 &=& v_1+2v_3+2v_4=-v_1-2v_2
 \nonumber
  \end{eqnarray} 
If $u=\alpha_1v_1+\alpha_2v_2+\alpha_4v_4$ is eigenvector of $A$ , $A\,u=\lambda\,u$, one easily finds :
 \begin{eqnarray}
&&\lambda=1 \quad , \quad u_1=v_1+v_4 \qquad , \nonumber\\
&&\lambda=\sqrt{5} \quad , \quad u_2=
\frac{3+\sqrt{5}}{4}v_1+ v_2 +
\frac{1-\sqrt{5}}{4}v_4
\qquad , \nonumber\\
&&\lambda=-\sqrt{5} \quad , \quad u_3= 
\frac{3-\sqrt{5}}{4}v_1+ v_2
+\frac{1+\sqrt{5}}{4}v_4
 \qquad  \nonumber\\
 \nonumber
  \end{eqnarray} 
To compare this solution with the general solution given in eqs. ($C.1$) and ($C.2$), one rewrites the eigenvectors in the form :\\
\begin{eqnarray}
&&\lambda=1 \quad , \quad u_1= \sum_{j=1}^4\cos \frac{\pi(2j-1)}{4}v_j \sim v_1-v_2-v_3+v_4 \qquad , \qquad\nonumber\\
&&\lambda=1+4 \cos \frac{2\pi}{5}=\sqrt{5} \quad , \quad u_2=\sum_{j=1}^4 \sin \frac{2\pi j}{5} v_j \sim
\sqrt{5+\sqrt{5}}\,(v_1-v_4)+
\sqrt{5-\sqrt{5}}\,(v_2-v_3)
\qquad , \nonumber\\
&&\lambda=1+4 \cos \frac{4\pi}{5}=-\sqrt{5} \quad , \quad u_3=\sum_{j=1}^4 \sin \frac{4\pi j}{5} v_j \sim
\sqrt{5-\sqrt{5}}\,(v_1-v_4)+
\sqrt{5+\sqrt{5}}\,(v_3-v_2)
\qquad  \nonumber\\
 \nonumber
  \end{eqnarray} 

  Finally we translate the eigenvectors (\ref{t.8}) into linear combinations of traces of products of $5$ matrix fields by first adding one unit, then taking the inverse in each permutation.
 We outline the derivation for the element $v_2=(2,4)P\left( (e-(1,4)\right)$.
$P$ is the sum of the $6$ permutations over the elements $\{1,2,3\}$. We manifest it with the notation $P_{S_{3}}(1,2,3)$. Furthermore $(2,4)P_{S_{3}}(1,2,3)=P_{S_{3}}(1,3,4)(2,4)$ and $(2,4)(1,4)=(1,4)(1,2)$. Then
$$v_2=P_{S_3}(1,3,4)(2,4)-P_{S_3}(1,3,4)(1,2)$$ 
According to the relations in Footnote 1 we add one unit in the symbols, evaluate the inverse of the elements and rewrite in terms of matrix fields
$$v_2 =  \sum_{p \in S_{3}} \Tr {\phi_{\underline{1}}\phi_2\phi_{\underline{5}}\phi_3 \phi_{4} }-\Tr 
{\phi_{\underline{1}}\phi_3\phi_{\underline{2}}\phi_4 \phi_{5} }$$ 
  In a similar way we obtain the remaining basis vectors $v_j$. To exhibit the symmetry in  compact equations, we denote underlined numbers as symbols which are fixed in the sum over permutations
  \footnote{  For example 
   \begin{eqnarray}
   v_3 &=&  \sum_{p \in S_3} \Tr { \phi_{ \underline{1}}\phi_2 \phi_3 \phi_{\underline{5}} \phi_4-
\phi_{ \underline{1}} \phi_3 \phi_4 \phi_{\underline{2}}\phi_5 }=
\nonumber\\ 
 &=& \Tr { \phi_1 
(\phi_2\phi_3\phi_5\phi_4+\phi_2\phi_4\phi_5\phi_3+\phi_3\phi_2\phi_5\phi_4 +\phi_3\phi_4\phi_5\phi_2+\phi_4\phi_2\phi_5\phi_3+\phi_4\phi_3\phi_5\phi_2)} \nonumber\\
&-&\Tr {\phi_1(\phi_3\phi_4\phi_2\phi_5+\phi_3\phi_5\phi_2\phi_4+\phi_4\phi_3\phi_2\phi_5+\phi_4\phi_5\phi_2\phi_3+
\phi_5\phi_3\phi_2\phi_4+\phi_5\phi_4\phi_2\phi_3) }
\nonumber
   \end{eqnarray} }.
  
 \begin{eqnarray}
v_1&=& \sum_{p \in S_3} \Tr {\phi_{\underline{1}}\phi_{\underline{5}}\phi_2\phi_3\phi_4-\phi_{\underline{1}}\phi_{\underline{2}}\phi_3\phi_4\phi_5} \quad , \nonumber\\
v_2 &=&  \sum_{p \in S_3}\Tr { \phi_{ \underline{1}}\phi_2 \phi_{\underline{5}}\phi_3\phi_4-
\phi_{ \underline{1}}\phi_3\phi_{\underline{2}}\phi_4\phi_5 }\quad , 
\nonumber\\
v_3 &=&  \sum_{p \in S_3}\Tr { \phi_{ \underline{1}}\phi_2\phi_3\phi_{\underline{5}}\phi_4-
\phi_{ \underline{1}}\phi_3\phi_4\phi_{\underline{2}}\phi_5 }\quad , 
 \nonumber\\ 
 v_4&=&\sum_{p \in S_3} \Tr { \phi_{ \underline{1}}\phi_2\phi_3\phi_4\phi_{\underline{5}}-
\phi_{ \underline{1}}\phi_3\phi_4\phi_5\phi_{\underline{2}} }
 \quad  \label{t.7}
\end{eqnarray}

\section{The replacements}

In the literature it is more usual to represent states of scalar fields in term of $3$ complex fields $Z$ , $W$ , $Y$ , rather then their $6$ real components $\phi_\alpha$. One may choose
\begin{eqnarray}
&&Z=\frac{1}{\sqrt{2}}(\phi_1+i\phi_2) \quad , \quad
W=\frac{1}{\sqrt{2}}(\phi_3+i\phi_4) \quad , \quad
Y=\frac{1}{\sqrt{2}}(\phi_5+i\phi_6) \quad , \nonumber\\
&&{\bar Z}=\frac{1}{\sqrt{2}}(\phi_1-i\phi_2) \quad , \quad
{\bar W}=\frac{1}{\sqrt{2}}(\phi_3-i\phi_4) \quad , \quad
{\bar Y}=\frac{1}{\sqrt{2}}(\phi_5-i\phi_6) \quad  \label{t.20}
\end{eqnarray}

For $\bigtriangleup_0 \leq 6$ our auxiliary model does not differ from superconformal Yang Mills theory,  in the sector we studied, then all the eigenstates we described, like in eqs.(\ref{t.3a}), (\ref{t.3b}) for $\bigtriangleup_0=4$
or eq.(\ref{t.7}) for $\bigtriangleup_0=5$, are eigenstates of the superconformal Yang Mills theory. Writing them as real matrix fields or, through eq.(\ref{t.20}) in terms of complex matrix fields is irrelevant.\\

However the relevance of our auxiliary model depends on the possibility of yielding the spectrum of superconformal Yang Mills theory for arbitrary values of $\bigtriangleup_0$. This is performed by a trivial replacement rule.\\
Given an eigenvector of the auxiliary model pertinent to the group $S_n$ , that is $\bigtriangleup_0=n+1$, it is a linear combination of basis vectors $\Tr {\phi_1\phi_{\alpha_2} \cdots \phi_{\alpha_{n+1}} }$. One can partition the set of flavours $\{1, \alpha_2 , \cdots , \alpha_{n+1}\}$ in $3$ sets , say of cardinality $n_1, n_2, n_3$ such that $n_1+n_2+n_3=n+1$ and each $n_j \geq 0$. Then replace for each term of the linear combination
all $\phi_{\alpha_j}$ of the first set with $Z$, those of the second set with $W$ , those of the third set with $Y$.
In this way one obtains a state linear combination of states all in the sector $\Tr {Z^{n_1}W^{n_2}Y^{n_3} }$.
This linear combination is an eigenstate of the  superconformal Yang Mills theory. \\

 This trivial replacement rule is correct because the trace operator $\sum_{l=1}^{n+1} K_{l,l+1}$ vanishes on states of the sector 
$\Tr {Z^{n_1}W^{n_2}Y^{n_3} } $  and the remaining part of the dilatation operator, the exchange operator $\sum P_{l,l+1}$, eq.(\ref{t.1}), acts in the same way irrespective of the replacement.\\

Let us illustrate this vanishing for a simple example. Let us consider a state $ \Tr { \cdots Z^m \cdots }$ where all couples of adjacent matrix fields before and after $Z^m$ have different flavours. Then the possible non-vanishing contributions of the action of the trace operator is due to its action on the $m-1$ pairs of adjacent $m$ matrix fields $Z$. If the first field $Z$ occurs at position $r$ inside the trace
\begin{eqnarray}
&&\sum_{l=1}^{n+1} K_{l,l+1}\, \Tr { \cdots Z^m \cdots }=\sum_{l=r}^{r+m-1} K_{l,l+1}\, \Tr { \cdots Z^m \cdots }
 \nonumber\\
&&= K_{r,r+1}\,\Tr { \cdots (Z^2)Z^{m-2} \cdots }+ K_{r+1,r+2}\,\Tr { \cdots Z(Z^2)Z^{m-3} \cdots }+ \cdots +
 \nonumber\\
&&+K_{r+m-1,r+m}\,\Tr { \cdots Z^{m-2}(Z^2) \cdots }
\nonumber
\end{eqnarray}\\
where we have exhibited inside parenthesis the pair of adjacent $Z$ fields at the sites where the trace operator acts.
Each term in the sum vanishes. Indeed for a term where the pair of $Z$ fields are on sites $(l,l+1)$ we have
$$K_{l,l+1}\, \Tr { \cdots (Z^2) \cdots }= K_{l,l+1}\, \Tr { \cdots (\phi_1^2-\phi_2^2+i\phi_1\phi_2+i\phi_2\phi_1) \cdots }=0$$

The different choices of fields in the sets $n_j$ may lead  to  inequivalent sets of eigenvectors for the same sector $\Tr {Z^{n_1}W^{n_2}Y^{n_3}}$.\\
 We  illustrate how the replacement rule applies to the basis vectors
(\ref{t.7}) with the choice $n_1=n_2=2$ , $n_3=1$, that is
 the sector $\Tr {Z^2W^2Y}$  \\

{\bf The sector $\Tr{ Z^2W^2Y}$.}\\
In the irreducible representation $(3,1)$ of the group $S_4$ each basis vector $v_j$ vanishes if the field $\phi_2$ and $\phi_5$ are replaced with the same complex field. The general solution ($C.1$) and ($C.2$) indicates one eigenvector with positive parity  and eigenvalue $\bigtriangleup_2=4$ and two eigenvectors with negative parity and eigenvalues $\bigtriangleup_2=5 \pm \sqrt{5}$.  We do two inequivalent  replacings : first $\{ \phi_1\,,\,\phi_2\}\to W$ , $\{ \phi_3\,,\,\phi_4\}\to Z$ , $\phi_5 \to Y$ and second
 $\{ \phi_2\,,\,\phi_3\}\to W$ , $\{ \phi_4\,,\,\phi_5\}\to Z$ , $\phi_1 \to Y$. These different replacements originate sets of  eigenvectors where the positive parity eigenvectors are inequivalent.
 As we know all the eigenvectors in the sector $\Tr {Z^2W^2Y}$ from direct diagonalization, see Appendix A, further different  replacements would not originate new eigenvectors.\\
The first replacement obtains the basis vectors
\begin{eqnarray}
v_1 &=& \Tr {(WZ^2W+ZWZW-ZW^2Z-W^2Z^2)Y} \quad , \nonumber\\
v_2 &=& \Tr {(-WZ^2W-ZWZW +ZW^2Z+Z^2W^2)Y}\quad , \nonumber\\
v_3 &=& \Tr {(-WZ^2W-WZWZ +ZW^2Z+W^2Z^2)Y}  \quad , \nonumber\\
v_4 &=& \Tr {(WZ^2W+WZWZ-ZW^2Z-Z^2W^2)Y}
 \quad  \nonumber
\end{eqnarray} 
They originate one positive parity eigenstate
\begin{eqnarray}
&&\bigtriangleup_2=4 \qquad , \qquad u=\sum_{j=0}^4 \cos \frac{(2j-1)\pi}{4} v_j\sim v_1-v_2-v_3+v_4= \quad \nonumber \\
&&\qquad =\Tr {(2WZ^2W-2ZW^2Z+ZWZW+WZWZ-W^2Z^2-Z^2W^2)Y}
\quad  \nonumber
\end{eqnarray} 
and two negative parity eigenstates
\begin{eqnarray}
&&\bigtriangleup_2=5-\sqrt{5} \qquad , \qquad u=\sum_{j=0}^4 \sin\frac{2\pi j}{5}v_j \sim \qquad \nonumber\\
&&\sim {\rm tr}\,[(ZWZW-WZWZ)Y]+(\sqrt{5}+2) {\rm tr}\,[
(Z^2W^2-W^2Z^2)Y] \qquad , \nonumber\\
&&\bigtriangleup_2=5+\sqrt{5} \qquad , \qquad u=\sum_{j=0}^4 \sin\frac{4\pi j}{5}v_j \sim \qquad \nonumber\\
&&\sim -(\sqrt{5}+2) {\rm tr}\,[(ZWZW-WZWZ)Y]+ {\rm tr}\,[
(Z^2W^2-W^2Z^2)Y] \qquad  
\quad  \nonumber
\end{eqnarray} 

The second replacement obtains the basis vectors
\begin{eqnarray}
v_1 &=& \Tr {(Z^2W^2-WZWZ-WZ^2W+ZWZW+ZW^2Z-W^2Z^2)Y} \quad , \nonumber\\
v_2 &=& \Tr {(-W^2Z^2+WZWZ+WZ^2W-ZWZW -ZW^2Z+Z^2W^2)Y}\quad , \nonumber\\
v_3 &=& \Tr {(-Z^2W^2+ZWZW+WZ^2W-WZWZ -ZW^2Z+W^2Z^2)Y}  \quad , \nonumber\\
v_4 &=& \Tr {(W^2Z^2-ZWZW-WZ^2W+WZWZ+ZW^2Z-Z^2W^2)Y}
 \quad  \nonumber
\end{eqnarray} 
They originate one positive parity eigenstate
\begin{eqnarray}
&&\bigtriangleup_2=4 \qquad , \qquad u=\Tr {(ZW^2Z-WZ^2W)Y}
\quad  \nonumber
\end{eqnarray} 
and two negative parity eigenstates equal to the previous ones.\\

  Anyway the substitutions here performed are just examples, other types of substitutions are possible and lead to further eigenstates in sectors with different charges. 
Indeed it is possible to replace just a subset of matrix fields $\phi_\alpha$ and rewrite the remaining ones in term of their defining complex fields.
For instance from the state $u_4$, given in eq.(\ref{t.3b})
  \begin{eqnarray}
u_4=\Tr { (\phi_1 \phi_2-\phi_2\phi_1)(\phi_3\phi_4-\phi_4\phi_3)} \quad  \nonumber
\end{eqnarray} 
 we may replace 
 $$(\phi_1 \phi_2-\phi_2\phi_1)=i({\bar W}W-W {\bar W}) \quad , \quad \phi_3 \to Z \quad , \quad \phi_4 \to Y $$
 and obtain the new eigenstate $\Tr{({\bar W}W-W {\bar W})(ZY-YZ)}$.\\

\section*{Appendix A. The lowest part of the spectrum}
Eigenstates of the dilatation operator are listed here with the main quantum number   $4 \leq \bigtriangleup_0 \leq 7$  ,  in order of increasing values of the one loop contribution
 \boldmath$\bigtriangleup_2$.\unboldmath \, The easiest way to obtain them is direct diagonalization of the dilatation operator in finite dimensional vector spaces invariant under its action (the sectors).\\

Since re-labelling of fields trivially leads to equivalent eigenstates, we only list eigenstates of the form $\Tr {Z^a W^b Y^c}$ with $a \geq b \geq c$.\\

Eigenstates with the same values of \boldmath$\bigtriangleup_0$ and
\boldmath$\bigtriangleup_2$ are here written as eigenstates of the parity operator. Occasionally further degeneracies occur and linear combinations of degenerate eigenstates with the same quantum numbers are equally valid.\\

  \boldmath$\bigtriangleup_0$\unboldmath$=4$ \quad , \quad 
 \boldmath$\bigtriangleup_2$\unboldmath$=0$ \, : 
$$ \Tr {Z^4} \quad , \quad \Tr {Z^3W} \quad , \quad 2\,\Tr {Z^2 W^2}+\Tr{ZWZW} \quad , \quad 
\Tr {Z^2(WY+YW)}+\Tr{ZWZY} \quad .$$\\
\boldmath$\bigtriangleup_0$\unboldmath$=4$ \quad , \quad 
 \boldmath$\bigtriangleup_2$\unboldmath$=4$ \, : \qquad
 $ \Tr{Z^2(WY-YW)} $\quad . \quad  \\ 
 
\boldmath$\bigtriangleup_0$\unboldmath$=4$ \quad , \quad 
 \boldmath$\bigtriangleup_2$\unboldmath$=6$ \, : 
$$\Tr{Z^2 W^2} -\Tr{ZWZW} \quad , \quad
\Tr{Z^2(WY+YW)}-2\,\Tr{ZWZY}
 \quad .$$\\ 

  \boldmath$\bigtriangleup_0$\unboldmath$=5$ \quad , \quad 
 \boldmath$\bigtriangleup_2$\unboldmath$=0$ \, :  
\begin{eqnarray}
&&\Tr {Z^5} \quad , \quad \Tr {Z^4W} \quad , \quad \Tr {Z^3 W^2+ Z^2WZW}\quad , \nonumber\\
&&\Tr {Z^3(WY+YW)} + \Tr{Z^2(WZY+YZW)} \quad , \quad \nonumber\\
&&\Tr{ Z^2(W^2Y+YW^2+WYW)+ZW(ZWY+ZYW+WZY)}  \quad .        
\nonumber
\end{eqnarray}\\
 
 \boldmath$\bigtriangleup_0$\unboldmath$=5$ \quad , \quad 
 \boldmath$\bigtriangleup_2$\unboldmath$=5-\sqrt{5}=8\sin^2 \frac{\pi}{5}$  \, :  
\begin{eqnarray}
&&\left(\sqrt{5}+1\right)\Tr{Z^3(YW-WY)}+2\,\Tr{Z^2(YZW-WZY)} \quad , \nonumber\\
&&\left(2+\sqrt{5}\right)\Tr{Z^2(YW^2-W^2Y)}+\Tr{ZWZ(YW-WY)} \quad .\nonumber
\end{eqnarray}\\

  \boldmath$\bigtriangleup_0$\unboldmath$=5$ \quad , \quad 
 \boldmath$\bigtriangleup_2$\unboldmath$=4$  \, :  
\begin{eqnarray}
&&\Tr{Z^3W^2-Z^2WZW}\quad , \quad \Tr{Z^2 WYW-W^2 ZYZ} \quad , \nonumber\\
&&\Tr{Z^2(WZY+YZW-ZWY-ZYW)} \quad , \nonumber\\
&&\Tr{ZWZ(YW+WY)-Z^2(W^2Y+YW^2) } \qquad .
\nonumber
\end{eqnarray}\\
 
 \boldmath$\bigtriangleup_0$\unboldmath$=5$ \quad , \quad 
 \boldmath$\bigtriangleup_2$\unboldmath$=6$  \, :  
$$\Tr{Z^2(W^2Y+YW^2-2WYW)+WZ(WYZ+WZY-2YZW)} \quad .$$

 \boldmath$\bigtriangleup_0$\unboldmath$=5$ \quad , \quad 
 \boldmath$\bigtriangleup_2$\unboldmath$=5+\sqrt{5}=8\sin^2 \frac{2\pi}{5}$  \, :  
\begin{eqnarray}
&&-\left(\sqrt{5}-1\right)\Tr{Z^3(YW-WY)}+2\,\Tr{Z^2(YZW-WZY)} \quad , \nonumber\\
&&\left(\sqrt{5}-2 \right)\Tr{Z^2(W^2Y-YW^2)}-\Tr{ZWZ(YW-WY)} \quad .
\nonumber
\end{eqnarray}\\

  \boldmath$\bigtriangleup_0$\unboldmath$=6$ \quad , \quad 
 \boldmath$\bigtriangleup_2$\unboldmath$=0$ \, :  
\begin{eqnarray}
&&\Tr {Z^6} \quad , \quad \Tr {Z^5W} \quad , \quad \Tr {2Z^4 W^2+2 Z^3WZW+Z^2WZ^2W}\quad , \nonumber\\
&& 3\, \Tr {Z^3W^3+W^2Z^2WZ+Z^2W^2ZW}+ \Tr {ZWZWZW} \qquad ,\nonumber\\
&& \Tr {Z^4(WY+YW)+Z^3(WZY+YZW)+Z^2WZ^2Y} \qquad ,\nonumber\\
&&{\rm tr}\, \left( (Z^3W^2+W^2Z^3+Z^2W^2Z+ZW^2Z^2+Z^2WZW+WZWZ^2+ZWZ^2W+WZ^2WZ+ \right.  \nonumber\\
&& \qquad \left. +WZ^3W+ZWZWZ)Y \right) \quad, \nonumber\\
&& 2 \,{\rm tr}\, \left(Z^2W^2Y^2+Y^2W^2Z^2+Z^2WYWY+YWYWZ^2+Z^2YW^2Y+W^2ZY^2Z+Z^2WY^2W+  \right.\nonumber\\
&& \quad +ZWZWY^2+Y^2WZWZ+ZWZYWY+W^2ZYZY+YZYZW^2+\nonumber\\
&& \quad \left. +WZWYZY+ZWYWZY \right)+  \Tr {WZYWZY+ZWYZWY} \quad .
\nonumber
\end{eqnarray}\\

  \boldmath$\bigtriangleup_0$\unboldmath$=6$ \quad , \quad 
 \boldmath$\bigtriangleup_2$\unboldmath$=2$ \, :  
\begin{eqnarray}
&& \Tr {Z^4WY-YWZ^4+Z^3WZY-YZWZ^3} \quad , \nonumber\\
&&-2 \,\Tr {(Z^3W^2-W^2Z^3)Y} -\Tr {(Z^2W^2Z-ZW^2Z^2)Y}- \Tr {(Z^2WZW-WZWZ^2)Y} 
\quad,\nonumber \\
&& -3\,\Tr {Z^2W^2Y^2-Y^2W^2Z^2}+  {\rm tr}\, \left(
Z^2YWYW-WYWYZ^2+Y^2WZWZ-ZWZWY^2+ \right.\quad  \nonumber\\
&& \qquad \left.+ZYZYW^2- W^2YZYZ\right) \qquad .
\nonumber
\end{eqnarray}\\

\boldmath$\bigtriangleup_0$\unboldmath$=6$ \quad , \quad 
 \boldmath$\bigtriangleup_2$\unboldmath$=5-\sqrt{5}=8\sin^2 \frac{\pi}{5}$  \, :  
\begin{eqnarray}
&&-(1+\sqrt{5})\,\Tr {Z^4W^2} +(\sqrt{5}-1)\,\Tr {Z^3WZW}+2 \Tr {Z^2WZ^2W}\qquad ,\nonumber\\
&&-2(2+\sqrt{5})\, \Tr {Z^3W^3}+(1+\sqrt{5})\, \Tr {Z^2W^2ZW+W^2Z^2ZW}+2\, \Tr {ZWZWZW} \quad , \nonumber\\
&&-(1+\sqrt{5}) \Tr {Z^4(WY+YW)}+(\sqrt{5}-1) \Tr {Z^3(WZY+YZW)}+4\, \Tr {Z^2WZ^2Y} \quad , \nonumber\\
&& 2\,\Tr {(Z^3W^2+W^2Z^3-Z^2W^2Z-ZW^2Z^2)Y}+ (\sqrt{5}-3)\, \Tr {(Z^2WZW+WZWZ^2)Y}+ \nonumber\\
&& \qquad +(5-3\sqrt{5})\, \Tr {ZWZWZY}+(1+\sqrt{5})\,\Tr {WZ^3WY}\quad ,\nonumber\\
&& -(\sqrt{5}+1)\, \Tr {(Z^2W^2Z+ZW^2Z^2)Y} +2\, \Tr {(ZWZ^2W+WZWZ^2)Y}+\nonumber\\
&& \qquad +(\sqrt{5}-3)\, \Tr {ZWZWZY}+(1+\sqrt{5})\, \Tr {WZ^3WY} \qquad ,\nonumber\\
&& 2\,\Tr {W^2ZYZY+YZYZW^2-Y^2ZWZW-WZWZY^2}+\nonumber\\
&& \qquad +(\sqrt{5}+1)\,\Tr {Z^2YW^2Y-Z^2WY^2W}+(\sqrt{5}-1)\,\Tr {ZWYWZY-ZWZYWY} \quad ,\nonumber\\
&& (\sqrt{5}-1)\, \Tr {WZYWZY+ZWYZWY}-2\sqrt{5}\, \Tr {W^2ZYZY+YZYZW^2}+\nonumber\\
&& \qquad +2\,\Tr {Z^2WYWY+YWYWZ^2+Y^2ZWZW+WZWZY^2-Z^2W^2Y^2-Y^2W^2Z^2}+\nonumber\\
&& \qquad +4\,\Tr {ZWZYWY-Z^2YW^2Y-W^2ZY^2Z}
+2(1+\sqrt{5})\,\Tr {Z^2WY^2W} \quad ,\nonumber\\
&& 2\,\Tr {W^2ZYZY+YZYZW^2-Z^2WYWY-YWYWZ^2}+\nonumber\\
&& \qquad +(\sqrt{5}-1)\,\Tr {WZWYZY-ZWZYWY}+(\sqrt{5}+1)\,\Tr {W^2ZY^2Z-Z^2WY^2W} \quad .
\nonumber
\end{eqnarray}\\

\boldmath$\bigtriangleup_0$\unboldmath$=6$ \quad , \quad 
 \boldmath$\bigtriangleup_2$\unboldmath$=7-\sqrt{13}$ \, :  
\begin{eqnarray}
&& 2(4+\sqrt{13})\,\Tr {Z^2W^2Y^2+Y^2W^2Z^2}-2(\sqrt{13}+3)\,\Tr {Z^2YW^2Y+W^2ZY^2Z+Y^2WZ^2W}+\nonumber\\
&& \qquad -2\,\Tr {Z^2WYWY+YWYWZ^2+Y^2ZWZW+WZWZY^2+W^2ZYZY+YZYZW^2}+\nonumber\\
&& \qquad +4\,\Tr {ZWZYWY+WZWYZY+ZWYWZY}+(\sqrt{13}+1)\,\Tr {WZYWZY+ZWYZWY} \,.
\nonumber
\end{eqnarray}\\

 \boldmath$\bigtriangleup_0$\unboldmath$=6$ \quad , \quad 
 \boldmath$\bigtriangleup_2$\unboldmath$=4$ \, :  $\Tr {Z^3(WZY-YZW)}$ \\
 
 \boldmath$\bigtriangleup_0$\unboldmath$=6$ \quad , \quad 
 \boldmath$\bigtriangleup_2$\unboldmath$=5$ \, :  
\begin{eqnarray}
&& \Tr {(Z^3W^2+W^2Z^3-Z^2W^2Z-ZW^2Z^2+Z^2WZW+WZWZ^2)Y}-2\,\Tr {WZ^3WY}\quad , \nonumber\\
&& \Tr {(Z^3W^2-W^2Z^3-Z^2W^2Z+ZW^2Z^2-Z^2WZW+WZWZ^2)Y} \quad , \nonumber\\
&& 2\, \Tr {Z^2WY^2W-Z^2YW^2Y+ZWYWZY-ZWZYWY }+\nonumber\\
&& \qquad +\Tr {W^2ZYZY+YZYZW^2-ZWZWY^2-Y^2WZWZ } \quad , \nonumber\\
&& 2\, \Tr {Z^2WY^2W-W^2ZYY^2Z+ WZWYZY-ZWZYWY }+\nonumber\\
&& \qquad +\Tr {W^2ZYZY+YZYZW^2-Z^2WYWY-YWYWZ^2 } \quad , \nonumber\\
&& \Tr {Z^2WYWY-YWYWZ^2+W^2ZYZY-YZYZW^2} \quad , \nonumber\\
&& \Tr {Z^2WYWY-YWYWZ^2+Y^2WZWZ-ZWZWY^2} \qquad .
\nonumber
\end{eqnarray}\\

\boldmath$\bigtriangleup_0$\unboldmath$=6$ \quad , \quad 
 \boldmath$\bigtriangleup_2$\unboldmath$=6$ \, :  
\begin{eqnarray}
&& \Tr {Z^2W^2ZW-W^2Z^2WZ } \quad , \quad  \Tr {Z^4WY-YWZ^4-Z^3WZY+YZWZ^3} \quad , \nonumber\\
&& \Tr {(ZWZ^2W-WZ^2WZ)Y} \quad,  \quad
 \Tr {(Z^2W^2Z-ZW^2Z^2-Z^2WZW+WZWZ^2)Y} \quad, \nonumber\\
&& {\rm tr}\, \left(Z^2W^2Y^2-Y^2W^2Z^2+Z^2YWYW-WYWYZ^2+Y^2WZWZ-ZWZWY^2+ \right.\quad , \nonumber\\
&& \qquad \left.+ZYZYW^2- W^2YZYZ\right) \qquad , \qquad \Tr {WZYWZY-ZWYZWY} \qquad .
\nonumber
\end{eqnarray}\\

\boldmath$\bigtriangleup_0$\unboldmath$=6$ \quad , \quad 
 \boldmath$\bigtriangleup_2$\unboldmath$=5+\sqrt{5}$ \, :  
\begin{eqnarray}
&&(\sqrt{5}-1)\,\Tr {Z^4W^2} -(\sqrt{5}+1)\,\Tr {Z^3WZW}+2 \Tr {Z^2WZ^2W} \qquad ,\nonumber\\
&&-2(2-\sqrt{5})\, \Tr {Z^3W^3}+(1-\sqrt{5})\, \Tr {Z^2W^2ZW+W^2Z^2ZW}+2\, \Tr {ZWZWZW} \quad , \nonumber\\
&&(\sqrt{5}-1)\, \Tr {Z^4(WY+YW)}-(\sqrt{5}+1)\, \Tr {Z^3(WZY+YZW)}+4\,\Tr {Z^2WZ^2Y} \quad , \nonumber\\
&& 2\,\Tr {(Z^3W^2+W^2Z^3-Z^2W^2Z-ZW^2Z^2)Y}- (\sqrt{5}+3)\, \Tr {(Z^2WZW+WZWZ^2)Y}+ \nonumber\\
&& \qquad +(5+3\sqrt{5})\, \Tr {ZWZWZY}+(1-\sqrt{5})\,\Tr {WZ^3WY}\quad ,\nonumber\\
&& (\sqrt{5}-1)\, \Tr {(Z^2W^2Z+ZW^2Z^2)Y} +2\, \Tr {(ZWZ^2W+WZWZ^2)Y}+\nonumber\\
&& \qquad -(\sqrt{5}+3)\, \Tr {ZWZWZY}+(1-\sqrt{5})\, \Tr {WZ^3WY} \qquad ,\nonumber\\
&& 2\,\Tr {ZWZWY^2+Y^2WZWZ-W^2ZYZY-YZYZW^2}+\nonumber\\
&& \qquad -(\sqrt{5}-1)\,\Tr {Z^2YW^2Y-Z^2WY^2W}+ (\sqrt{5}+1)\, \Tr {ZWYWZY-ZWZYWY} \quad ,\nonumber\\
&& (1+\sqrt{5})\, \Tr {WZYWZY+ZWYZWY}-2\sqrt{5}\, \Tr {W^2ZYZY+YZYZW^2}+\nonumber\\
&& \qquad -2\,\Tr {Z^2WYWY+YWYWZ^2+Y^2ZWZW+WZWZY^2-Z^2W^2Y^2-Y^2W^2Z^2}+\nonumber\\
&& \qquad -4\,\Tr {ZWZYWY-Z^2YW^2Y-W^2ZY^2Z}
+2(\sqrt{5}-1)\,\Tr {Z^2WY^2W} \quad ,\nonumber\\
&& -2\,\Tr {W^2ZYZY+YZYZW^2-Z^2WYWY-YWYWZ^2}+\nonumber\\
&& \qquad +(\sqrt{5}+1)\,\Tr {WZWYZY-ZWZYWY}+(\sqrt{5}-1)\,\Tr {W^2ZY^2Z-Z^2WY^2W} \quad .
\nonumber
\end{eqnarray}\\

\boldmath$\bigtriangleup_0$\unboldmath$=6$ \quad , \quad 
 \boldmath$\bigtriangleup_2$\unboldmath$=7+\sqrt{13}$ \, :  
\begin{eqnarray}
&& 2(4-\sqrt{13})\,\Tr {Z^2W^2Y^2+Y^2W^2Z^2}+2(\sqrt{13}-3)\,\Tr {Z^2YW^2Y+W^2ZY^2Z+Y^2WZ^2W}+\nonumber\\
&& \qquad -2\,\Tr {Z^2WYWY+YWYWZ^2+Y^2ZWZW+WZWZY^2+W^2ZYZY+YZYZW^2}+\nonumber\\
&& \qquad +4\,\Tr {ZWZYWY+WZWYZY+ZWYWZY}-(\sqrt{13}-1)\,\Tr {WZYWZY+ZWYZWY} \,.
\nonumber
\end{eqnarray}\\

\boldmath$\bigtriangleup_0$\unboldmath$=7$ \quad , \quad Sector $\{Z^5 W^2\}$ , eigenstates with positive parity :\\
\noindent
\boldmath$\bigtriangleup_2$\unboldmath$=0$ \quad , \quad $u= \Tr {Z^5 W^2+Z^4WZW+Z^3WZ^2W}$ \quad , \\
\boldmath$\bigtriangleup_2$\unboldmath$=2$ \quad , \quad $u=\Tr {Z^5 W^2-Z^3WZ^2W}$ \quad ,\\
\boldmath$\bigtriangleup_2$\unboldmath$=6$ \quad , \quad $u=\Tr {Z^5 W^2}- 2\,\Tr {Z^4WZW}+\Tr{Z^3WZ^2W}$\\

\boldmath$\bigtriangleup_0$\unboldmath$=7$ \quad , \quad Sector $\{Z^5 WY\}$ , eigenstates with positive parity :\\
\noindent
\boldmath$\bigtriangleup_2$\unboldmath$=0$ \quad , \quad $u=\sum_{j=1}^6 \Tr {Z^{j-1}YZ^{6-j}W}$ \quad , \\
\boldmath$\bigtriangleup_2$\unboldmath$=2$ \quad , \quad $u=\Tr {Z^5(WY+YW)-Z^3(WZ^2Y+YZ^2W)}$ \quad ,\\
\boldmath$\bigtriangleup_2$\unboldmath$=6$ \quad , \quad $u=\Tr {Z^5(WY+YW)}- 2\,\Tr {Z^4(WZY+YZW)}+\Tr{Z^3(WZ^2Y+YZ^2W)}$\\

\boldmath$\bigtriangleup_0$\unboldmath$=7$ \quad , \quad Sector $\{Z^5 WY\}$ , eigenstates with negative parity :\\
\noindent
\boldmath$\bigtriangleup_2$\unboldmath$=8 \sin^2 (\pi/7) $ \quad , \quad $u= \sin(2\pi/7)\,\Tr {Z^5(WY-YW)}+
 \sin(4\pi/7)\,\Tr {Z^4(WZY-YZW)}+$
 ${ }\qquad + \sin(6\pi/7)\,\Tr {Z^3(WZ^2Y-YZ^2W)}$\quad ,\\
\boldmath$\bigtriangleup_2$\unboldmath$=8 \sin^2 (2\pi/7) $ \quad , \quad $u= -\sin(4\pi/7)\,\Tr {Z^5(WY-YW)}+
 \sin(6\pi/7)\,\Tr {Z^4(WZY-YZW)}+$
 ${ }\qquad + \sin(2\pi/7)\,\Tr {Z^3(WZ^2Y-YZ^2W)}$\quad ,\\
\boldmath$\bigtriangleup_2$\unboldmath$=8 \sin^2 (3\pi/7) $ \quad , \quad $u= \sin(6\pi/7)\,\Tr {Z^5(WY-YW)}-
 \sin(2\pi/7)\,\Tr {Z^4(WZY-YZW)}+$
 ${ }\qquad + \sin(4\pi/7)\,\Tr {Z^3(WZ^2Y-YZ^2W)}$\quad .\\

\boldmath$\bigtriangleup_0$\unboldmath$=7$ \quad , \quad Sector $\{Z^4W^3\}$ , eigenstates with positive parity :\\
\noindent
\boldmath$\bigtriangleup_2$\unboldmath$=0$ \quad , \quad $u=\Tr {Z^4W^3+Z^3W^2ZW+W^2Z^3WZ+Z^2W^2Z^2W+Z^2WZWZW}$\\
\boldmath$\bigtriangleup_2$\unboldmath$=2$ \quad , \quad $u=\Tr {-2\,Z^4W^3+Z^2W^2Z^2W+Z^2WZWZW}$\\
\boldmath$\bigtriangleup_2$\unboldmath$=5$ \quad , \quad $u=\Tr {2\,Z^4W^3-3(Z^3W^2Z W+W^2Z^3WZ)+2\,Z^2W^2Z^2W+2\,Z^2WZWZW}$\\
\boldmath$\bigtriangleup_2$\unboldmath$=6$ \quad , \quad $u=\Tr {-Z^2W^2Z^2W+Z^2WZWZW}$\\

\boldmath$\bigtriangleup_0$\unboldmath$=7$ \quad , \quad Sector $\{Z^4W^3\}$ , eigenstates with negative parity :\\
\noindent
\boldmath$\bigtriangleup_2$\unboldmath$=5$ \quad , \quad $u=\Tr {Z^3W^2ZW-W^2Z^3WZ}$\\

\boldmath$\bigtriangleup_0$\unboldmath$=7$ \quad , \quad Sector $\{Z^4W^2Y\}$.
Basis vectors with positive parity : \\
$v_1= \Tr {(Z^4W^2+W^2Z^4)Y }$ , $v_2=\Tr {(Z^3W^2Z+ZW^2Z^3)Y}$ , $v_3=\Tr {(Z^3WZW+WZWZ^3)Y }$, $v_4=\Tr {Z^2W^2Z^2Y}$ , $v_5= \Tr {(Z^2WZWZ+ZWZWZ^2)Y}$ , $v_6= \Tr {(Z^2WZ^2W+WZ^2WZ^2)Y}$, $v_7=\Tr {ZWZ^2WZY}$ , $v_8=\Tr {(ZWZ^3W+WZ^3WZ)Y}$ , $v_9=\Tr {WZ^4WY}$ .\\
\noindent
Eigenvalues and eigenstates : \\
\boldmath$\bigtriangleup_2$\unboldmath$=0$ \quad , \quad $u=\sum_1^9 v_j$ \quad , \\
\boldmath$\bigtriangleup_2$\unboldmath$=2$ \quad , \quad  $u=2(v_1+v_9)-(v_4+v_5+v_6+v_7)$ \quad , \\
\boldmath$\bigtriangleup_2$\unboldmath$=2$ \quad , \quad  $u=-(v_1+v_2+v_4)+(v_6+v_7+v_8)$ \quad , \\
\boldmath$\bigtriangleup_2$\unboldmath$=4$ \quad , \quad  $u=-v_1-v_3-v_6+v_8+2(v_4+v_9)$ \quad , \\
\boldmath$\bigtriangleup_2$\unboldmath$=5$ \quad , \quad  $u=2(v_1+v_4+v_5+v_6+v_7+v_9)-3(v_2+v_3+v_8)$ \quad , \\
\boldmath$\bigtriangleup_2$\unboldmath$=6$ \quad , \quad  $u=-v_1+v_2+v_5-v_8-2(v_4-v_9)$ \quad , \\
\boldmath$\bigtriangleup_2$\unboldmath$=6$ \quad , \quad  $u=v_2-v_3-v_4+v_6$ \quad ,\\
\boldmath$\bigtriangleup_2$\unboldmath$=6$ \quad , \quad  $u=v_2-v_3-v_5+v_6$ \quad , \\
\boldmath$\bigtriangleup_2$\unboldmath$=8$ \quad , \quad  $u=-(v_1+v_5)+3(v_3-v_8)+2(v_4+v_9)-4v_5+8v_7$ \quad , \\

\boldmath$\bigtriangleup_0$\unboldmath$=7$ \quad , \quad Sector $\{Z^4W^2Y\}$.
Basis vectors with negative parity : \\
$v_1= \Tr {(Z^4W^2-W^2Z^4)Y }$ , $v_2=\Tr {(Z^3W^2Z-ZW^2Z^3)Y}$ , $v_3=\Tr {(Z^3WZW-WZWZ^3)Y }$,  $v_4= \Tr {(Z^2WZWZ-ZWZWZ^2)Y}$ , $v_5= \Tr {(Z^2WZ^2W-WZ^2WZ^2)Y}$ , $v_6=\Tr {(ZWZ^3W-WZ^3WZ)Y}$ .\\
\noindent
Eigenvalues and eigenstates : \\
\boldmath$\bigtriangleup_2$\unboldmath$=8 \sin^2 \pi/7$ \quad , \quad $u=\sin (2\pi/7) \left(v_1+v_3+v_5+v_6 \right)+
\sin (4\pi/7) \left(v_1+v_2+v_4-v_6 \right)+$ \nonumber\\
$ \qquad +\sin (6\pi/7) \left(v_2+v_3-v_4-v_5 \right)$ \quad ,\\
\boldmath$\bigtriangleup_2$\unboldmath$=4$ \quad , \quad $u=-2v_1+v_2+v_3+v_4+v_5$ \quad ,\\
\boldmath$\bigtriangleup_2$\unboldmath$=8 \sin^2 2\pi/7$ \quad , \quad $u=\sin (4\pi/7) \left(v_1+v_3+v_5+v_6 \right)+
\sin (8\pi/7) \left(v_1+v_2+v_4-v_6 \right)+$ \nonumber\\
$ \qquad +\sin (12\pi/7) \left(v_2+v_3-v_4-v_5 \right)$ \quad ,\\
\boldmath$\bigtriangleup_2$\unboldmath$=5$ \quad , \quad $u=v_2-v_3+v_6$ \quad ,\\
\boldmath$\bigtriangleup_2$\unboldmath$=8 \sin^2 3\pi/7$ \quad , \quad $u=\sin  (6\pi/7) \left(v_1+v_3+v_5+v_6 \right)+
\sin (12\pi/7) \left(v_1+v_2+v_4-v_6 \right)+$ \nonumber\\
$ \qquad +\sin (4\pi/7) \left(v_2+v_3-v_4-v_5 \right)$ \quad ,\\
\boldmath$\bigtriangleup_2$\unboldmath$=8$ \quad , \quad $u=-v_2+v_3+3(v_4-v_5)+2v_6$ \quad \\

\boldmath$\bigtriangleup_0$\unboldmath$=7$ \quad , \quad Sector $\{Z^3W^3Y\}$.
Basis vectors with positive parity : \\
$v_1=\Tr {Z^3W^3Y+YW^3Z^3} \quad , \quad v_2=\Tr {Z^2W^3ZY+YZW^3Z^2} \quad , $\\
$ v_3=\Tr {Z^2W^2ZWY+YWZW^2Z^2} \quad , \quad v_4= \Tr {Z^2WZW^2Y+YW^2ZWZ^2} \quad , \quad $\\
  $v_5=\Tr {ZW^2ZWZY+YZWZW^2Z} \quad , \quad v_6=\Tr {ZW^2Z^2WY+YWZ^2W^2Z} \quad ,$\\
  $ v_7=\Tr {ZWZ^2W^2Y+YW^2Z^2WZ} \quad , \quad v_8=\Tr {ZWZWZWY+YWZWZWZ} \quad  ,$\\
$ v_9=\Tr {W^2Z^3WY+YWZ^3W^2} \quad , \quad v_{10}=\Tr {WZ^2WZWY+YWZWZ^2W} \quad .$\\
\noindent
Eigenvalues and eigenstates : \\
\boldmath$\bigtriangleup_2$\unboldmath$=0$ \quad , \quad $u=\sum_1^{10} v_j$ \quad , \\
\boldmath$\bigtriangleup_2$\unboldmath$=2$ \quad , \quad  $u=2(v_1+v_9)-(v_3+v_5+v_6+v_8)$ \quad , \\
\boldmath$\bigtriangleup_2$\unboldmath$=2$ \quad , \quad  $u=2(v_1+v_2)-(v_6+v_7+v_8+v_{10})$ \quad , \\
\boldmath$\bigtriangleup_2$\unboldmath$=4$ \quad , \quad  $u=v_1-v_2+v_4-v_9$ \quad , \\
\boldmath$\bigtriangleup_2$\unboldmath$=5$ \quad , \quad  $u=v_2+v_7+v_{10}-2(v_1+v_6+v_8)-3v_4$ \quad , \\
\boldmath$\bigtriangleup_2$\unboldmath$=5$ \quad , \quad  $u=v_3+v_5+v_9-2(v_1+v_6+v_8)-3v_4$ \quad , \\
\boldmath$\bigtriangleup_2$\unboldmath$=6$ \quad , \quad  $u=v_3-v_5$ \quad ,\\
\boldmath$\bigtriangleup_2$\unboldmath$=6$ \quad , \quad  $u=v_6-v_8$ \quad , \\
\boldmath$\bigtriangleup_2$\unboldmath$=6$ \quad , \quad  $u=v_7-v_{10}$ \quad , \\
\boldmath$\bigtriangleup_2$\unboldmath$=8$ \quad , \quad  $u=v_1-v_2-v_9+2(v_3+v_5-v_6+v_7-v_8+v_{10})-3v_4$ \quad . \\

\boldmath$\bigtriangleup_0$\unboldmath$=7$ \quad , \quad Sector $\{Z^3W^3Y\}$.
Basis vectors with negative parity : \\
$v_1=\Tr {Z^3W^3Y-YW^3Z^3} \quad , \quad v_2=\Tr {Z^2W^3ZY-YZW^3Z^2} \quad , $\\
$ v_3=\Tr {Z^2W^2ZWY-YWZW^2Z^2} \quad , \quad v_4= \Tr {Z^2WZW^2Y-YW^2ZWZ^2} \quad , \quad $\\
  $v_5=\Tr {ZW^2ZWZY-YZWZW^2Z} \quad , \quad v_6=\Tr {ZW^2Z^2WY-YWZ^2W^2Z} \quad ,$\\
  $ v_7=\Tr {ZWZ^2W^2Y-YW^2Z^2WZ} \quad , \quad v_8=\Tr {ZWZWZWY-YWZWZWZ} \quad  ,$\\
$ v_9=\Tr {W^2Z^3WY-YWZ^3W^2} \quad , \quad v_{10}=\Tr {WZ^2WZWY-YWZWZ^2W} \quad .$\\
Eigenvalues and eigenstates : \\
\boldmath$\bigtriangleup_2$\unboldmath$=8\,\sin^2 \pi/7$ \quad , \quad  $u= \left(  \sin 2\pi/7+\sin 4\pi/7+\sin 6\pi/7 \right)v_1+
(\sin 4\pi/7)(v_2-v_9)+(\sin 2\pi/7)(v_3+v_7)+
\left(\sin 2\pi/7+\sin 4\pi/7-\sin 6\pi/7 \right)v_4+(\sin 6\pi/7)(-v_5+v_{10})+ \left(\sin 2\pi/7-\sin 4\pi/7-\sin 6\pi/7\right)v_6+
\left(\sin 2\pi/7-\sin 4\pi/7+\sin 6\pi/7\right)v_8$ \quad ,\\
\boldmath$\bigtriangleup_2$\unboldmath$=4$ \quad , \quad  $u=v_2+v_3-v_7+v_9$ \quad , \\
\boldmath$\bigtriangleup_2$\unboldmath$=8\,\sin^2 2\pi/7$ \quad , \quad  $u=  
\left(  \sin 2\pi/7-\sin 4\pi/7+\sin 6\pi/7 \right)v_1+
(\sin 6\pi/7)(v_2-v_9)-(\sin 4\pi/7)(v_3+v_7)+
\left(-\sin 2\pi/7-\sin 4\pi/7+\sin 6\pi/7 \right)v_4+(\sin 2\pi/7)(-v_5+v_{10})- \left(\sin 2\pi/7+\sin 4\pi/7+\sin 6\pi/7\right)v_6+
\left(\sin 2\pi/7-\sin 4\pi/7-\sin 6\pi/7\right)v_8$ \quad ,\\
\boldmath$\bigtriangleup_2$\unboldmath$=5$ \quad , \quad  $u=v_2-v_4+v_7-v_{10}$ \quad , \\
\boldmath$\bigtriangleup_2$\unboldmath$=5$ \quad , \quad  $u=v_3-v_4+v_5-v_9$ \quad , \\
\boldmath$\bigtriangleup_2$\unboldmath$=8\,\sin^2 3\pi/7$ \quad , \quad  $u=\left( -\sin 2\pi/7+\sin 4\pi/7+\sin 6\pi/7\right)v_1+\left(\sin 2\pi/7 \right)\left(v_9-v_2 \right)+(\sin 6\pi/7)\left(v_3+v_7\right)-\left(\sin2\pi/7+\sin 4\pi/7-\sin 6\pi/7 \right)v_4+(\sin 4\pi/7) (-v_5+v_{10})+$\\
$+ \left(\sin 2\pi/7-\sin 4\pi/7+\sin 6\pi/7\right)v_6+
\left(\sin 2\pi/7+\sin 4\pi/7+\sin 6\pi/7\right)v_8$ \quad ,\\
\boldmath$\bigtriangleup_2$\unboldmath$=8$ \quad , \quad  $u=v_2-v_3+v_7+v_9+2(v_5+v_{10})$ \quad , \\
The remaining $3$ eigenvectors correspond to the eigenvalues \boldmath$\bigtriangleup_2$\unboldmath$=7-\lambda$ where $\lambda$ are the roots of the equation : $\lambda^3-\lambda^2-17 \lambda+25=0$.\\

\boldmath$\bigtriangleup_0$\unboldmath$=7$ \quad , \quad Sector $\{Z^3W^2Y^2\}$.
Basis vectors with positive parity : \\
$v_1=\Tr {Z^3W^2Y^2+Y^2W^2Z^3} \quad , \quad v_2=\Tr {Z^2W^2ZY^2+Y^2 ZW^2Z^2} \quad , $\\
$ v_3=\Tr {Z^2W ZWY^2+Y^2WZW Z^2} \quad , \quad v_4= \Tr {ZWZ^2WY^2+Y^2WZ^2WZ} \quad , \quad $\\
  $v_5=\Tr {ZWZWZY^2} \quad , \quad v_6=\Tr {WZ^3WY^2} \quad ,$\\
  $ v_7=\Tr {Z^3WYWY+YWYW Z^3} \quad , \quad v_8=\Tr {Z^2WZYWY+YWYZWZ^2} \quad  ,$\\
$ v_9=\Tr {Z^2W^2YZY+YZYW^2Z^2} \quad , \quad v_{10}=\Tr {ZWZWYZY+YZYWZWZ} \quad ,$\\
$v_{11}=\Tr {ZW^2ZYZY} \quad , \quad v_{12}=\Tr {WZ^2WYZY} \quad , \quad v_{13}=\Tr {Z^3YW^2Y} \quad ,$\\
$v_{14}=\Tr {Z^2WYWZY+YZWYWZ^2} \quad , \quad v_{15}=\Tr {Z^2WYZWY+YWZYWZ^2} \quad ,$\\
$v_{16}=\Tr {ZWZYWZY+YZWYZWZ} \quad , \quad v_{17}=\Tr {ZW^2YZ^2Y+YZ^2YW^2Z} \quad,$\\
$v_{18}=\Tr {WZWYZ^2Y} \quad.$\\
Eigenvalues and eigenstates : \\
\boldmath$\bigtriangleup_2$\unboldmath$=0$ \quad , \quad  $u=\sum_1^{18} v_j \quad ,$\\
\boldmath$\bigtriangleup_2$\unboldmath$=2$ \quad , \quad  $u=-2v_1+4v_2+v_3+3v_5-4(v_6+v_7)-v_8+v_9+2v_{10}+3v_{11}+v_{12}-4v_{13}-v_{14}+v_{18} \quad ,$\\
\boldmath$\bigtriangleup_2$\unboldmath$=2$ \quad , \quad  $u=v_1-2v_2-v_3-v_4-2v_5+v_6+2v_7-v_{10}-v_{11}-v_{12}+3v_{13}+v_{14}+v_{17} \quad ,$\\
\boldmath$\bigtriangleup_2$\unboldmath$=2$ \quad , \quad  $u=-3v_2-v_3-2v_5+2v_6+2v_7+v_8-v_9-v_{10}-2v_{11}+2v_{13}+v_{14}+v_{15}+v_{16} \quad , $\\
\boldmath$\bigtriangleup_2$\unboldmath$=4$ \quad , \quad  $u=v_1-2v_2-v_3+v_4+2v_6-v_7-2v_8+v_9+v_{10}+2v_{12}-2v_{13}-v_{16}+v_{17}+2v_{18} \quad ,$\\
\boldmath$\bigtriangleup_2$\unboldmath$=4$ \quad , \quad  $u=-v_3+2v_6-v_8+v_9-2v_{13}+v_{14} \quad ,$\\
\boldmath$\bigtriangleup_2$\unboldmath$=5$ \quad , \quad  $u=v_1+v_2-3v_4-2v_5+4v_6+v_7-3v_9+v_{10}+4v_{11}-2v_{12}-2v_{13}-3v_{14}+v_{15}+v_{16}+4v_{18} \quad ,$\\
\boldmath$\bigtriangleup_2$\unboldmath$=5$ \quad , \quad  $u=-v_1-v_2+2v_3+v_4-2v_5-v_7+2v_8+v_9-v_{10}-2v_{12}-2v_{13}+v_{14}-v_{15}-v_{16}+2v_{17}\quad ,$\\
\boldmath$\bigtriangleup_2$\unboldmath$=6$ \quad , \quad  $u=v_5-v_{11}-v_{12}+v_{18} \quad ,$\\
\boldmath$\bigtriangleup_2$\unboldmath$=6$ \quad , \quad  $u=v_1-2v_3+v_4+2v_5-v_6+v_8-v_9-v_{10}+v_{11}-v_{12}-v_{13}+v_{17} \quad ,$\\
\boldmath$\bigtriangleup_2$\unboldmath$=6$ \quad , \quad  $u=v_3+
v_6-v_7-v_8-v_{11}-v_{12}+v_{13}+v_{16} \quad ,$\\
\boldmath$\bigtriangleup_2$\unboldmath$=6$ \quad , \quad  $u=v_2+
v_3-2v_5+v_6-v_7-v_8-v_{10}-v_{11}+v_{12}+v_{13}+v_{15} \quad ,$\\
\boldmath$\bigtriangleup_2$\unboldmath$=6$ \quad , \quad  $u=v_3-v_8-v_9+v_{14} \quad ,$\\
\boldmath$\bigtriangleup_2$\unboldmath$=8$ \quad , \quad  $u=-2v_1+
4v_2+v_3+2v_4-4v_5-6v_6+10v_7-5v_8-v_9-2v_{10}-12v_{11}+8v_{12}-10v_{13}-7v_{14}-8v_{15}+10(v_{16}+v_{17}) \quad ,$\\
\boldmath$\bigtriangleup_2$\unboldmath$=8$ \quad , \quad  $u=-4v_1+8v_2-3v_3+24v_4-28v_5-22v_6+20v_7-15v_8+3v_9-4(v_{10}+v_{11}+v_{12})-10 v_{13}-9v_{14}-16 v_{15}+20(v_{16}+v_{18})$ \quad . \\
The remaining $3$ eigenvectors correspond to the eigenvalues 
\boldmath$\bigtriangleup_2$\unboldmath$=7-\lambda$ where $\lambda$ are the roots of the equation : $\lambda^3+\lambda^2-17\lambda-25=0$.\\

\boldmath$\bigtriangleup_0$\unboldmath$=7$ \quad , \quad Sector $\{Z^3W^2Y^2\}$.
Basis vectors with negative parity : \\
$v_1=\Tr {Z^3W^2Y^2-Y^2W^2Z^3} \quad , \quad v_2=\Tr {Z^2W^2ZY^2-Y^2 ZW^2Z^2} \quad , $\\
$ v_3=\Tr {Z^2W ZWY^2-Y^2WZW Z^2} \quad , \quad v_4= \Tr {ZWZ^2WY^2-Y^2WZ^2WZ} \quad , \quad $\\
  $ v_5=\Tr {Z^3WYWY-YWYW Z^3} \quad , \quad v_6=\Tr {Z^2WZYWY-YWYZWZ^2} \quad  ,$\\
$ v_7=\Tr {Z^2W^2YZY-YZYW^2Z^2} \quad , \quad v_{8}=\Tr {ZWZWYZY-YZYWZWZ} \quad ,$\\
 $v_{9}=\Tr {Z^2WYWZY-YZWYWZ^2} \quad , \quad v_{10}=\Tr {Z^2WYZWY-YWZYWZ^2} \quad ,$\\
$v_{11}=\Tr {ZWZYWZY-YZWYZWZ} \quad , \quad v_{12}=\Tr {ZW^2YZ^2Y-YZ^2YW^2Z} \quad.$\\
 Eigenvalues and eigenstates : \\
 \boldmath$\bigtriangleup_2$\unboldmath$=8 \sin^2 \pi/7 $ \quad , \quad  $u=(2 \sin 2\pi/7 +2 \sin 4\pi/7+\sin 6\pi/7)v_1+
 (\sin 2\pi/7+\sin 4\pi/7)v_2+(\sin 2\pi/7+\sin 4\pi/7+\sin 6\pi/7)(v_3+v_7)+ (\sin 6\pi/7 )(v_4+v_{12})+(2\sin 4\pi/7-\sin 6\pi/7)v_5+(\sin 2\pi/7-\sin 4\pi/7+2\sin 6\pi/7)(v_6+v_9)+(-\sin 2\pi/7+3\sin 4\pi/7-3\sin 6\pi/7)v_8+(-\sin 2\pi/7+\sin 4\pi/7)v_{10}+(\sin 2\pi/7-\sin 4\pi/7+\sin 6\pi/7)v_{11} \quad , $\\
\boldmath$\bigtriangleup_2$\unboldmath$=4$ \quad , \quad  $u=-v_3-v_4+v_7+v_{12} \quad ,$\\
\boldmath$\bigtriangleup_2$\unboldmath$=4$ \quad , \quad  $u=-v_1+v_2-v_3-2v_4+v_5+v_6+v_7+v_9+v_{10} \quad ,$\\
\boldmath$\bigtriangleup_2$\unboldmath$=8 \sin^2 2\pi/7 $ \quad , \quad  $u=( \sin 2\pi/7 -2 \sin 4\pi/7+2\sin 6\pi/7)v_1+
 (\sin 6\pi/7-\sin 4\pi/7)v_2+(\sin 2\pi/7-\sin 4\pi/7+\sin 6\pi/7)(v_3+v_7)+ (\sin 2\pi/7 )(v_4+v_{12})+(2\sin 6\pi/7-\sin 2\pi/7)v_5+(2\sin 2\pi/7-\sin 4\pi/7-\sin 6\pi/7)(v_6+v_9)+(\sin 4\pi/7-3\sin 2\pi/7+3\sin 6\pi/7)v_8+(\sin 4\pi/7+\sin 6\pi/7)v_{10}+(\sin 2\pi/7-\sin 4\pi/7-\sin 6\pi/7)v_{11} \quad , $\\
\boldmath$\bigtriangleup_2$\unboldmath$=5$ \quad , \quad  $u=v_3-v_4+v_6-v_7-v_9+v_{12} \quad ,$\\
\boldmath$\bigtriangleup_2$\unboldmath$=5$ \quad , \quad  $u=-v_1-v_2+v_4+v_5+v_7+v_8+v_9-v_{10}+v_{11} \quad ,$\\
\boldmath$\bigtriangleup_2$\unboldmath$=8 \sin^2 3\pi/7 $ \quad , \quad  $u=(-2 \sin 2\pi/7 +\sin 4\pi/7+2\sin 6\pi/7)v_1+
 (\sin 6\pi/7-\sin 2\pi/7)v_2+(\sin 4\pi/7-\sin 2\pi/7+\sin 6\pi/7)(v_3+v_7)+ (\sin 4\pi/7 )(v_4+v_{12})-(2\sin 2\pi/7+\sin 4\pi/7)v_5+(\sin 2\pi/7+2\sin 4\pi/7+\sin 6\pi/7)(v_6+v_9)-(3\sin 2\pi/7+3\sin 4\pi/7+\sin 6\pi/7)v_8-(\sin 2\pi/7+\sin 6\pi/7)v_{10}+(\sin 2\pi/7+\sin 4\pi/7+\sin 6\pi/7)v_{11} \quad , $\\
\boldmath$\bigtriangleup_2$\unboldmath$=8$ \quad , \quad  $u=v_3-v_4-2v_6-v_7+2v_9+v_{12} \quad ,$\\
\boldmath$\bigtriangleup_2$\unboldmath$=8$ \quad , \quad  $u=v_1+v_2-3v_3+2v_4-v_5+3v_6-v_7+2v_8-v_9+v_{10}+2v_{11} \quad ,$\\
The remaining $3$ eigenvectors correspond to the eigenvalues 
\boldmath$\bigtriangleup_2$\unboldmath$=7-\lambda$ where $\lambda$ are the roots of the  cubic equation : $\lambda^3-\lambda^2-17\lambda+25=0$ .\\

\section*{Appendix B. The Eigenvalues}

 The text \cite{enc}, provides a table with all the irreducible representation for the generators of the permutation group $S_n$ up to $n=7$. It is easy to obtain all the irreducible representations for the operator 
 ${\bf A}$ in eq.(\ref{t.3}), up to the same order. We  list here the representations of the operator {\bf A} corresponding to $S_n$, $3 \leq n \leq 6$, and the  eigenvalues.\\
 
To denote the irreducible representation, we write the sequence of integers corresponding to the number of boxes in the horizontal rows of the Young tableaux. For instance $(2^2,1^2)$ is the irreducible representation of $S_6$ where the Young tableau has two boxes in the first two horizontal rows, and one box in the third and fourth row.\\

 If $\bigtriangleup_0=4$ , the relevant group is $S_3$. Beside the two $1$-dimensional representations corresponding to the partitions $(3)$ and $(1^3)$  , 
we only need the $2$-dimensional representation corresponding to the partition $(2,1)$  which may be chosen
 \begin{eqnarray}
(2,1) \qquad , \qquad {\bf A}=\left(
\begin{array}{ccc}
-1 & 1\\
1 & -1 \end{array} \right) \quad , \quad \lambda=0 \quad , \quad  \lambda=-2 \nonumber
\end{eqnarray}
 If $\bigtriangleup_0=5$ , the group is $S_4$. Beside the two $1$-dimensional representations corresponding to the partitions $(4)$ and $(1^4)$  , there are : one  $2$-dimensional representation 
corresponding to the partition $(2^2)$ 
and two  $3$-dimensional representations corresponding to the partitions $(3,1)$ and $(2,1^2)$. They respectively 
 may be chosen
 \begin{eqnarray}
(2^2)\quad &,&{\bf A}=\left(
\begin{array}{ccc}
0 & -1\\
-1 & 0 \end{array} \right) \quad , \quad \lambda=\pm 1 \quad , \quad   \nonumber \\
(3,1) \quad &,&{\bf A}=\left(
\begin{array}{ccc}
-1 & 2 & -2\\
0 & 2 & -1 \\
-2 & 1 & 0\end{array} \right) \quad , \quad \lambda=\pm \sqrt{5} \quad , \quad   \lambda=1 \quad ,\nonumber \\
(2,1^2) \quad &,&{\bf A}=\left(
\begin{array}{ccc}
0 & -1 & 2\\
1 & -2 & 2 \\
2 & 0 & 1\end{array} \right) \quad , \quad \lambda=\pm \sqrt{5} \quad , \quad   \lambda=-1 \quad ,\nonumber 
\end{eqnarray}

If $\bigtriangleup_0=6$ the group is $S_5$. Beside the two $1$-dimensional representations corresponding to the partitions $(5)$ and $(1^5)$  , there are :  
 two  $4$-dimensional representations , corresponding to the partitions $(4,1)$ and $(2,1^3)$, 
two  $5$-dimensional representations corresponding to the partitions $(3,2)$ and $(2^2,1)$, 
and one $6$-dimensional representation corresponding to the partition $(3,1^2)$ . They respectively 
 may be chosen
 $$(4,1) \quad , \quad 
{\bf A}=\left(\begin{array}{cccc}
0 & 2 & 0 &-2\\
0 & 2 & 2 & -2 \\
-2 & 2 & 3 & -1\\
-2 & 0 & 1 & 1\end{array} \right) \quad , \quad \lambda=4 \quad , \quad \lambda=1\pm \sqrt{5} \quad , \quad   \lambda=0 \quad , $$
$$(2,1^3) \quad , \quad
{\bf A}=\left(\begin{array}{cccc}
-3 & -1 & 0 &0\\
-1 & -3 & 0 & 0 \\
0 & 0 & -2 & 0\\
0 & 0 & 0 & -2\end{array} \right) \quad , \quad \lambda=-4 \quad , \quad \lambda=-2 \quad {\rm three} \quad {\rm times} \quad , \quad  $$
$$(3,2) \quad , \quad
{\bf A}=\left(\begin{array}{ccccc}
0 & 1 & 0 & 0 & -1\\
0 & 1 & 0 & 0 & 0 \\
0 & 0 & 2 & 0 & -2\\
-1 & 1 & -1 & 1 & 1\\
-1 & 1 & -1 & 1 & 0\end{array} \right) \quad , \quad \lambda=1 \quad {\rm twice} \quad , \quad \lambda=1\pm \sqrt{5} \quad , \quad   \lambda=0 \quad ,$$
$$(2^2,1) \quad , \quad
{\bf A}=\left(\begin{array}{ccccc}
3 & 0 & -4 & 0 & 2\\
1 & 1 & -2 & 0 & 1 \\
1 & -1 & -4 & 2 & 1\\
-1 & 1 & 2 & -1 & 2\\
-1 & -1 & 2 & 2 & -3\end{array} \right) \quad , \quad \lambda=1 \quad {\rm twice} \quad , \quad \lambda=-1\pm \sqrt{13} \quad , \quad   \lambda=-4 \quad ,$$
$$(3,1^2) \quad , \quad
{\bf A}=\left(\begin{array}{cccccc}
0 & 0 & 1 & 0 & 1 & 0\\
2 & 0 & 1 & -2 & 2 & 1 \\
2 & 1 & 1 & -2 & 0 & 2\\
1 & 0 & 0 & -1 & 1 & 1\\
1 & 2 & 0 & -1 & 0 & 2\\
0 & 1 & 1 & 0 & 0 & 2\end{array} \right) \quad , \quad \lambda=1 \quad {\rm twice} \quad , \quad \lambda=-2
 \quad {\rm twice} \quad, \quad   \lambda=4 \quad , \quad \lambda=0  $$
 
To make easier the comparison with eigenvalues already evaluated,  see Table 3.2 in the  reference \cite{r1}, we remark that the Laplacian eigenvalues $\bigtriangleup_2=5\pm \sqrt{5}$ and $\bigtriangleup_2=7 \pm \sqrt{13}$ are the roots of the equations $E^2-10 E +20=0$ and $E^2-14 E+36=0$ respectively.\\

 If $\bigtriangleup_0=7$ , the group is $S_6$. Beside the two $1$-dimensional representations corresponding to the partitions $(6)$ and $(1^6)$  , there are :  
 four  $5$-dimensional representations  corresponding to the partitions $(5,1)$ , $(3^2)$ , $(2^3)$, $(2,1^4)$, 
 two  $9$-dimensional representations, corresponding to the partitions  $(4,2)$, $(2^2,1^4)$, 
two  $10$-dimensional representations, corresponding to the partitions  $(4,1^2)$, $(3,1^3)$, 
and one $16$-dimensional representation corresponding to the partition  $(3,2,1)$. They respectively may be chosen
$$(5,1) \quad ,\quad
{\bf A}=\left(\begin{array}{ccccc}
1 & 2 & 0 & 0 & -2\\
0 & 3 & 2 & 0 & -2 \\
-2 & 2 & 3 & 2 & -2\\
-2 & 0 & 2 & 4 & -1\\
-2 & 0 & 0 & 1 & 2\end{array} \right)\quad , \quad \lambda=1 \quad , \quad \lambda=5, \quad \lambda^3-7\lambda^2+7\lambda+7=0 \quad ,  $$
$$(3^2) \quad , \quad
{\bf A}=\left(\begin{array}{ccccc}
2 & -1 & 1 & 1 & -3\\
0 & 2 & 1 & 0 & -2 \\
0 & 0 & 3 & 0 & -2\\
0 & -1 & 0 & 3 & -1\\
1 & -3 & 2 & 3 & -5\end{array} \right)\quad , \quad \lambda=2 \quad  {\rm twice}\quad , \quad 
\lambda^3-\lambda^2-17 \lambda+25=0 \quad ,$$
$$(2^3) \quad , \quad
{\bf A}=\left(\begin{array}{ccccc}
4 & -1 & -2 & 2 & 3\\
3 & -3 & 0 & 0 & 2 \\
2 & 0 & -3 & 1 & 1\\
3 & -1 & 0 & -2 & 2\\
1 & 0 & 0 & 0 & -1\end{array} \right)\quad , \quad \lambda=-2 \quad  {\rm twice}\quad , \quad 
\lambda^3+\lambda^2-17 \lambda-25=0 \quad , $$
$$(2,1^4) \quad , \quad
{\bf A}=\left(\begin{array}{ccccc}
-2 & -1 & 2 & -2 & 2\\
1 & -4 & 2 & 0 & 0 \\
0 & 2 & -3 & 2 & 0\\
0 & 0 & 2 & -3 & 2\\
2 & -2 & 2 & 0 & -1\end{array} \right)\quad , \quad \lambda=-1 \quad , \quad \lambda=-5, \quad \lambda^3+7\lambda^2+7\lambda-7=0 \quad , $$
 $$(4,2) \quad , \quad
{\bf A}=\left(\begin{array}{ccccccccc}
1 & 0 & 1 & 1 & 0 & -1  & 0 & -1 & 0\\
0 & 0 & 1 & 1 & 1 & -1 & 1 & -1 & -1 \\
0 & -1 & 3 & 0 & 1 & 0 & 1 & -2 & -1\\
-1 & 1 & 0 & 2 & 1 & -1 & 0 & 1 & -1\\
-1 & 0 & 1 & 1 & 2 & 0 & 0 & -1 & 0\\
0 & -1 & 1 & -1 & 1 & 3 & 1 & -1 & -2\\
-1 & 0 & 0 & 1 & 0 & -1 & 2 & 1 & 0\\
-1 & -1 & 0 & 2 & 0 & -1 & 1 & 1 & 1\\
0 & -1 & 0 & 0 & 1 & -1 & 1 & 0 & 1
\end{array} \right)\quad , \quad \lambda=-1 \quad {\rm twice}\quad , \quad \lambda=1 \quad {\rm twice}\quad ,$$
 $$\quad \lambda=2 \quad {\rm twice}\quad , \quad
\lambda=3 \quad {\rm twice}\quad , \quad 
\lambda=5, $$
$$(2^2,1^2) \quad , \quad
{\bf A}=\left(\begin{array}{ccccccccc}
0 & 0 & 0 & -2 & -1 & 0  & 1 & 0 & 1\\
0 & 0 & 0 & 1 & -2 & 0 & 2 & -1 & 0 \\
-1 & 1 & -2 & 1 & 0 & 0 & 0 & 0 & -1\\
-1 & 1 & -1 & -3 & 0 & 1 & -1 & 0 & 0\\
-1 & 0 & 0 & 1 & -2 & 1 & 0 & 0 & 0\\
0 & -2 & 1 & 1 & 1 & -2 & 0 & 1 & 1\\
0 & 0 & 0 & -1 & 1 & 0 & -3 & 1 & -1\\
1 & -1 & 0 & -1 & 0 & 1 & 0 & -1 & 1\\
0 & 1 & -1 & 0 & -1 & 1 & 0 & 0 & -2
\end{array} \right)\quad , \quad \lambda=-1 \quad {\rm twice}\quad , \quad $$
 $$\qquad \lambda=1 \quad {\rm twice}\quad ,\quad \lambda=-2 \quad {\rm twice}\quad , \quad
\lambda=-3 \quad {\rm twice}\quad , \quad 
\lambda=-5, \qquad $$
$$(4,1^2) \quad , \quad
{\bf A}=\left(\begin{array}{ccccccccccc}
1 & 0 & 1 & 0 & 0 & 0  & 0 & 1 & 0 & 0\\
2 & 0 & 1 & 0 & 1 & 0 & -1 & 1 & 1 & 0 \\
2 & 1 & 1 & -1 & 1 & 1 & -1 & 0 & 1 & 0\\
1 & 1 & 0 & 1 & 1 & 0 & -2 & 1 & 1 & 1\\
1 & 2 & 1 & 0 & 2 & 1 & -2 & -1 & 2 & 1\\
0 & 1 & 1 & 0 & 1 & 2 & -1 & -1 & 0 & 2\\
1 & 0 & 0 & 0 & 0 & 0 & 0 & 1 & 0 & 1\\
1 & 1 & 0 & 1 & 0 & 0 & -1 & 1 & 1 & 1\\
0 & 1 & 0 & 1 & 1 & 0 & -1 & 0 & 1 & 2\\
0 & 0 & 0 & 1 & 0 & 1 & 0 & 0 & 0 & 3
\end{array} \right)\quad , \quad \lambda=-1 \quad {\rm twice}\quad , \quad \lambda=0 \quad {\rm twice}\quad ,$$
 $$\quad \lambda=1  \quad , \quad
\lambda=3 \quad {\rm twice}\quad , \quad 
\lambda^3-7\lambda^2+7\lambda+7=0, $$
$$(3,1^3) \quad , \quad
{\bf A}=\left(\begin{array}{ccccccccccc}
-3 & 2 & -1 & 1 & -2 & 1  & -1 & 0 & 0 & 0\\
0 & -1 & 1 & 0 & 0 & -2 & 1 & 1 & -1 & 0 \\
0 & 0 & -1 & 1 & 1 & -1 & -1 & 0 & 1 & -1\\
0 & 1 & -1 & 0 & -1 & 2 & -2 & -1 & 1 & 0\\
-1 & 0 & 0 & 0 & -2 & 1 & 0 & -1 & 0 & 0\\
0 & -1 & 0 & 0 & 1 & -2 & 1 & 1 & -1 & 0\\
-1 & 1 & -1 & 0 & 0 & 0 & -1 & 1 & 0 & 0\\
0 & 0 & 0 & 0 & -1 & 1 & 0 & -1 & 1 & -1\\
0 & -1 & 1 & 0 & 1 & -2 & 1 & 1 & 0 & 0\\
0 & 0 & -1 & 1 & 0 & 1 & -1 & -2 & 2 & -1
\end{array} \right)\quad , \quad \lambda=-3 \quad {\rm twice}\quad , \quad $$
$$ \lambda=0 \quad {\rm twice}\quad ,\quad \lambda=-1  \quad , \quad
\lambda=1 \quad {\rm twice}\quad , \quad 
\lambda^3+7\lambda^2+7\lambda-7=0,  $$
$$(3,2,1) \quad , \quad
{\bf A}=\left(\begin{array}{cccccccccccccccc}
2& -1& 1& -1& 0& -1& 2& 0& -2& 2& -2& 0& -1& 1& -1& 1\\
2& -1& -2& 2&  0& -2& 1& 1& -1& 1& 0& 1& -3& 1& -2& 1\\
2& -1& 0& 1& 2& 0& 0& 1& -2&    0& -3& 1& 1& -1& 0& 3\\
1& 1& 0& 1& 0& 1& 0& 1& -1& 0& -2& 1& -1& 0& -1&     3\\
0& 0& 2& -1& 2& -1& 1& 1& -1& 0& -3& 0& 0& 0& 0& 2\\
0& 0& 1& 1& 0& 0& -1& 2& -1& 0& -1& 0& 0& 0& 0& 2\\
2& -1& -1& 0& 1& -1& 3& 1& -1&     1& -2& 1& -2& 0& -2& 3\\
1& 1& -1& 0& 1& 1& 0& 2& 0& -1& -2& 2&     0& -1& -1& 2\\
1& -1& 0& 0& 0& 0& 0& 0& -2& 2& 0& 0& 0& 0& 1& 0\\
1&     0& -1& 1& -1& 0& 0& 0& 1& 0& 0& 2& -1& 0& 0& 0\\
0& 0& 1& -1& 0& 0& 0& 0&     0& 1& -3& 0& 1& 0& 0& 1\\
0& 0& 1& 0& -1& 1& -1& 0& 0& 1& 0& -1& 1& 1&   2& 0\\
1& -1& 0& 1& 0& -1& 0& 0& -1& 0& 0& 1& -1& 1& 0& 1\\
1& 0& 0& 1&     0& 1& -1& 0& -1& -1& 0& 1& 2& 0& 2& 1\\
0& 0& 0& 0& -1& 0& 0& -1& 1& 0&     1& 1& 0& 1& 0& -1\\
0& 0& 1& 0& -1& 1& 0& -1& -1& 0& 2& -1& 0& 2& 2& -2
\end{array} \right)\quad , \quad
  $$
$\lambda=3$ twice, $\lambda=-3$ twice, $\lambda=1$ three times, 
$\lambda=-1$ three times, $\lambda^3-\lambda^2-17\lambda+25=0$,
$\lambda^3+\lambda^2-17\lambda-25=0$.\\

 To make easier the comparison with eigenvalues already evaluated,  \cite{be} , \cite{mz}, \cite{be2}, 
 we remark that the $3$ roots of $\lambda^3-7\lambda^2+7\lambda+7=0$ are $\lambda=3-4 \cos[(2p+1)\pi/7]$ with $p=0,1,2$. This is easily proved by extracting the real part of the identity $\sum_{k=0}^6 e^{2\pi i k/7}=0$ then obtaining a cubic equation for $\cos(\pi/7)$.
 The $3$ roots of $\lambda^3+7\lambda^2+7\lambda-7=0$ are the opposite of the previous roots.\\
The $3$ roots of the cubic equation $\lambda^3-\lambda^2-17 \lambda+25=0$ are mapped into $3$ eigenvalues $\bigtriangleup_2$ of the Laplacian by $E=7-\lambda$ and are the roots of the equation $E^3-20 E^2+116 E-200=0$, which is quoted in Table 3.4 of the  reference
\cite{r1}. In the same way, the roots of the equation $\lambda^3+\lambda^2-17 \lambda-25=0$ , which are opposite of the previous ones, become roots of the equation $E^3-22 E^2+144 E-248=0$.\\

\section*{Appendix C.  The Eigenvectors}
We are interested in the eigenstates corresponding to some irreducible representation of $S_n$.
We generalize the method used in Sect.3 for the representation $(3,1)$ of $S_4$, of degree $f=3$. Here we find basis vectors and eigenvectors for the two irreducible representations of $S_n$, the representation $(n-1,1)$ and its conjugate $(2,1^{n-2})$, both of degree $f=n-1$.\\
Only the first one provides important sequences of eigenvectors of the dilatation operator of superconformal Yang Mills theory, after proper replacements.\\

{\bf The representation $(n-1,1)$ of the group $S_n$.}\\

For any representation of the permutation group $S_n$ of degree $f$, one obtains a matrix of order $f$ which represents the operator ${\bf A}$ in eq.(\ref{t.3}). As the degree of the representation increases, the evaluations becomes more cumbersome. However there are two irreducible representations which may be easily derived for arbitrary $S_n$ : they are of degree $n-1$ and are pair-conjugate.\\
\epsfig{file=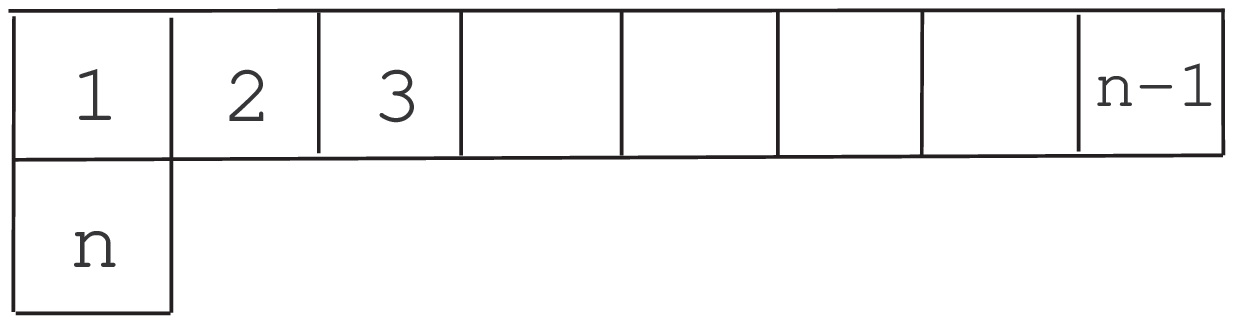 ,height=3cm}\\
The Young tableau  has $n-1$ boxes in the first row and one in the second row. \\

The set of elements in $S_n$ is partitioned in $n$ sets $a_j$, $j=1,2,..,n$. Each $a_j$ contains $(n-1)!$ elements and they may be written in the form $(j,n) \cdot g$ with $g \in S_{n-1}$.\\
The Young projector operator $Y$ associated to the irreducible representation of $S_n$ corresponding to the partition $(n-1,1)$ is $Y=PQ$ , where $P$ is the sum of the $(n-1)!$ permutations of $S_{n-1}$ and $Q=e-(1,n)$.\\
We choose $n$ basis vectors $v_j$ :
$$ v_j=(j,n) \cdot PQ \quad , \quad {\rm for} \, j=1,2..,n-1 \quad , \quad v_n=e \cdot PQ $$
Since $\sum_1^n v_j=0$ , one basis vector may be eliminated, but equations will be more neat by keeping all $n$ vectors.\\
Any permutation in $S_n$ which multiplies from left a basis vector $v_j$ obtains a basis vector. For example :\\
$(j,j+1)v_j=v_{j+1}$ , $(j,j+1)v_{j+1}=v_j$ , $(1,2,3,..,n)v_j=v_{j+1}$ , $(n , n-1,..,2,1)v_j=v_{j-1}$ \\

The action of the operator ${\bf A}=(1,2)+(2,3)+(3,4)+..+(n-1,n)+(1,2,...,n)+(n,n-1,..,2,1)$ on each basis vector is given by the system

\begin{eqnarray}
 A v_1&=&  (n-2)v_1+2v_2+v_n \quad , \nonumber\\
 Av_2 &=& 2v_1+(n-3)v_2+2v_3 \quad , \nonumber\\
  Av_3 &=& 2v_2+(n-3)v_3+2v_4 \quad , \nonumber\\
..&=&...   ... \quad , \nonumber\\
Av_{n-1} &=&2v_{n-2}+(n-3)v_{n-1}+2v_n \quad , \nonumber\\
A v_n &=& v_1+2v_{n-1}+(n-2)v_n \quad
  \nonumber
  \end{eqnarray} 
 If $u=\sum_1^n \alpha_j v_j$ is eigenvector of $A$ , $A\,u=\lambda\,u$, the set of coefficients $\alpha_j$ is determined by the system
 \begin{eqnarray}
    (n-2)\alpha_1+2\alpha_2+\alpha_n &=& \lambda \alpha_1\quad , \nonumber\\
  2\alpha_1+(n-3)\alpha_2+2\alpha_3 &=& \lambda \alpha_2\quad , \nonumber\\
  2\alpha_2+(n-3)\alpha_3+2\alpha_4 &=& \lambda \alpha_3 \quad , \nonumber\\
..&=&...   ... \quad , \nonumber\\
2\alpha_{n-2}+(n-3)\alpha_{n-1}+2\alpha_n &=&\lambda \alpha_{n-1}\quad , \nonumber\\
\alpha_1+2\alpha_{n-1}+(n-2)\alpha_n  &=& \lambda \alpha_n \quad
  \nonumber
  \end{eqnarray} 
   
  The $n-1$
 eigenvectors   are found by solving linear recurrence relations.  They form two sequences : 
 \begin{eqnarray}
\lambda_k=n-3+4 \cos 2k\pi/(n+1) &,& u^{(k)}=\sum_{j=1}^n \sin \left( \frac{2k\pi j}{n+1}\right) v_j \quad , \quad \nonumber\\
k=1,2,..,k_{max} &,& k_{max}<(n+1)/2 \quad , \quad {\rm and} \nonumber \\
 \lambda_k=n-3+4 \cos 2k\pi/n  &,& 
 u^{(k)}=\sum_{j=1}^n \cos \left( \frac{k\pi(2j-1)}{n}\right) v_j \quad , \quad \nonumber\\
 (C.1) \qquad k=1,2,..,k_{max} &,& k_{max}<n/2 . 
\nonumber
   \end{eqnarray}

Since the group $S_n$ is associated with $\bigtriangleup_0=n+1$, the two sets of eigenvalues translate into two sets for the one loop contribution to the dilatation dimension :\\
$\bigtriangleup_2=8 \sin^2 \frac{\pi k}{n+1}$ , $k=1,2,..,k_{max}$ , $k_{max}<(n+1)/2$ , and \\
$\bigtriangleup_2=8 \sin^2 \frac{\pi k}{n}$ , $k=1,2,..,k_{max}$ , $k_{max}<n/2$ . \\

We rewrite the basis vectors and the eigenvectors in term of traces of products of matrix fields by a straightforward generalization of the procedure leading from the basis vectors (\ref{t.8}) to their form (\ref{t.7})

 \begin{eqnarray}
\qquad v_1&=& \sum_{p \in S_{n-1}} \Tr {\phi_{\underline{1}}\phi_{\underline{n+1}}\phi_2\phi_3\cdots \phi_n-\phi_{\underline{1}}\phi_{\underline{2}}\phi_3\phi_4\cdots \phi_{n+1}} \quad , \nonumber\\
v_2 &=&  \sum_{p \in S_{n-1}} \Tr { \phi_{ \underline{1}}\phi_2 \phi_{\underline{n+1}}\phi_3\cdots\phi_n-
\phi_{ \underline{1}}\phi_3\phi_{\underline{2}}\phi_4\phi_5\cdots \phi_{n+1} }
\nonumber\\
v_3 &=&  \sum_{p \in S_{n-1}} \Tr { \phi_{ \underline{1}}\phi_2\phi_3\phi_{\underline{n+1}}\phi_4\cdots \phi_n-
\phi_{ \underline{1}}\phi_3\phi_4\phi_{\underline{2}}\phi_5\cdots \phi_{n+1}}
 \nonumber\\ 
 \cdots &=& \cdots  \quad \cdots \nonumber\\
 (C.2) \qquad v_n&=&\sum_{p \in S_{n-1}} \Tr {\phi_{ \underline{1}}\phi_2\phi_3\cdots \phi_n\phi_{\underline{n+1}}-
\phi_{ \underline{1}}\phi_3\phi_4\phi_5 \cdots \phi_{n+1}\phi_{\underline{2}} }
\nonumber
\end{eqnarray} 
where we
  use the compact notation of footnote 2 to indicate the elements remaining fixed in the sum over permutations.
It is useful to notice that , for every $j$ , the states $v_j$ and $v_{n-j+1}$ are related by parity.\\

{\bf The representation $(2,1^{n-2})$ of $S_n$.}\\

Although it will turn out to be of more limited use, we  derive a set of basis vectors and the eigenvectors which correspond to the irreducible representation conjugate to the previous one. Actually only slight changes occur in the derivation.\\

 \epsfig{file=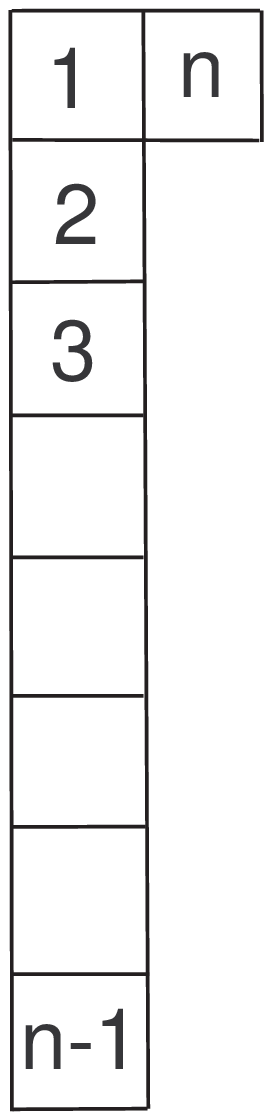,height=6cm}
We define $Q$ the sum of all permutations of the group $S_{n-1}$  with their sign, that is $(+1)$ or $(-1)$ depending on the even or odd numbers of transposition that relate the permutation to the identity permutation. 
$$Q=\sum_{p \in S_{n-1}} ({\rm sign}\,p)\,p $$
One easily finds that $-Q+\sum_{j=1}^{n-1} (j,n)Q$ is the opposite of the sum of all the  $n!$ permutations of the group $S_n$, with their sign.\\
We may choose a set of $n$ basis vectors
$$z_j=(j,n)Q\left(e+(1,n)\right) \quad ,\quad {\rm for}\quad j=1,2,\cdots ,n-1 \quad , \quad z_n=-Q\left(e+(1,n)\right) $$
Then $\sum_1^n z_n=0$. and one easily obtains the action of any transposition on the basis vectors :
\begin{eqnarray}
&&(j,j+1)(j,n)Q=(j+1,n)(j,j+1)Q=-(j+1,n)Q \quad , \quad {\rm then}\quad \nonumber\\
&&(j,j+1)z_j=-z_{j+1}\quad , \quad
(j,j+1)z_{j+1}=-z_j \quad
  \nonumber
  \end{eqnarray} 
 However $(1,2,\cdots ,n)z_j=-(-1)^n z_{j+1}$ and $(n,\cdots ,2,1)z_j=-(-1)^n z_{j-1}$.\\

 If $n$ is even we have
\begin{eqnarray}
 A z_1&=&  -(n-2)z_1-2z_2-z_n \quad , \nonumber\\
 Az_2 &=& -2z_1-(n-3)z_2-2z_3 \quad , \nonumber\\
  A z_3 &=& -2z_2-(n-3)z_3-2z_4 \quad , \nonumber\\
..&=&...   ... \quad , \nonumber\\
Az_{n-1} &=&-2z_{n-2}-(n-3)z_{n-1}-2z_n \quad , \nonumber\\
A z_n &=& -z_1-2z_{n-1}-(n-2)z_n \quad
  \nonumber
  \end{eqnarray} 
Eigenvalues of the operator $A$ are the opposite of the ones of the conjugate representation and eigenvectors are the same, where $z_j$ now replaces $v_j$, that is two sequences : \\
$\lambda_k=3-n-4 \cos 2k\pi/(n+1)$ , $u^{(k)}=\sum_{j=1}^n \sin \left( \frac{2k\pi j}{n+1}\right) z_j $ ,
$k=1,2,..,k_{max}$ , $k_{max}<(n+1)/2$ , and \\
 $\lambda_k=3-n-4 \cos 2k\pi/n$ , 
 $u^{(k)}=\sum_{j=1}^n \cos \left( \frac{k\pi(2j-1)}{n}\right) z_j $ ,
  $k=1,2,..,k_{max}$ , $k_{max}<n/2$ . \\

If $n$ is odd integer, we obtain
\begin{eqnarray}
 A z_1&=&  -(n-2)z_1+z_n \quad , \nonumber\\
..&=&...   ... \quad , \nonumber\\
 Az_j &=& -(n-3)z_j \quad , \nonumber\\
..&=&...   ... \quad , \nonumber\\
A z_n &=& z_1-(n-2)z_n \quad , \quad {\rm that}\quad {\rm is}
\nonumber\\
A&=&(3-n)I+
\left( \begin{array}{cccccccc}
  -1 & 0 & 0 & 0 & 0 & ... & 1\\
  0 & 0 & 0 & 0 & 0 & ... & 0\\
  0 & 0 & 0 & 0 & 0 &... & 0\\
  0 & 0 & 0 & 0 & 0 &... & 0\\
  ..& ..& .. & .. & ..& .. &..\\
  0 & 0 & 0 &  0 & ...     & 0 & 0\\ 
  1 & 0 & 0 & 0 & ... & 0 & -1
  \end{array} \right)
\nonumber
  \end{eqnarray} 
The $n$ eigenvalues of $A$ are :\\
$\lambda=1-n$ , $\lambda=3-n$ , $n-1$ times.\\
One of the degenerate eigenvalues $\lambda=3-n$ is associated to the null vector $\sum_1^n z_j$ then the $n-1$ eigenvalues of the irreducible representation, if $n$ is  odd integer, are:\\
$\lambda=1-n$ , $\lambda=3-n$ , $n-2$ times.\\

We proceed to rewrite the basis vectors $z_j$ and the eigenvectors in term of traces of products of matrix fields as we did for  the basis vectors $v_j$.\\
 
$Q$ is the sum of the $(n-1)!$ permutations over the elements $\{1,2,\cdots,n-1\}$. We manifest it with the notation $Q_{S_{n-1}}(1,2,\cdots,n-1)$. Furthermore $(2,n)Q_{S_{n-1}}(1,2,3,\cdots,n-1)=Q_{S_{n-1}}(1,n,3,\cdots,n-1)(2,n)$ and $(2,n)(1,n)=(1,n)(1,2)$. Then
$$z_2=Q_{S_{n-1}}(1,n,3,\cdots,n-1)(2,n)-Q_{S_{n-1}}(1,n,3,\cdots,n-1)(1,2)$$ 
According to the relations in Footnote 1 we add one unit in the symbols, evaluate the inverse of the elements and rewrite in terms of matrix fields. Some care is needed for the sign of the permutation.
We use the compact notation of footnote 2 to indicate the elements remaining fixed in the sum over permutations. Except for a overall factor $(-1)^n$ we find \\
 \begin{eqnarray}
z_1&=& (-1)^n \sum_{p \in S_{n-1}} ({\rm sign}\,p)\,  \Tr {\phi_{\underline{1}}\phi_{\underline{n+1}}\phi_2\phi_3\cdots \phi_n}+\sum_{p \in S_{n-1}} ({\rm sign}\,p)\,  \Tr {
\phi_{\underline{1}}\phi_{\underline{2}}\phi_3\phi_4\cdots \phi_{n+1}}\quad , \nonumber\\
z_2 &=& (-1)^{n+1} \sum_{p \in S_{n-1}}({\rm sign}\,p)\, \Tr { \phi_{ \underline{1}}\phi_2 \phi_{\underline{n+1}}\phi_3\cdots\phi_n}-
\sum_{p \in S_{n-1}}({\rm sign}\,p)\, \Tr {
\phi_{ \underline{1}}\phi_3\phi_{\underline{2}}\phi_4\phi_5\cdots \phi_{n+1} }
\nonumber\\
z_3 &=& (-1)^n \sum_{p \in S_{n-1}}({\rm sign}\,p)\,\Tr { \phi_{ \underline{1}}\phi_2\phi_3\phi_{\underline{n+1}}\phi_4\cdots \phi_n}+
\sum_{p \in S_{n-1}}({\rm sign}\,p)\,\Tr { 
\phi_{ \underline{1}}\phi_3\phi_4\phi_{\underline{2}}\phi_5\cdots \phi_{n+1}}
 \nonumber\\ 
 \cdots &=& \cdots  \quad \cdots \nonumber\\
 z_n&=&- \sum_{p \in S_{n-1}}({\rm sign}\,p)\,\Tr { \phi_{ \underline{1}}\phi_2\phi_3\cdots \phi_n\phi_{\underline{n+1}}}+
(-1)^{n+1}  
 \sum_{p \in S_{n-1}}({\rm sign}\,p)\,\Tr {
\phi_{ \underline{1}}\phi_3\phi_4\phi_5 \cdots \phi_{n+1}\phi_{\underline{2}}}
 \quad  \nonumber
\end{eqnarray} 

This irreducible representation provides eigenvalues and eigenvectors for the Super Yang Mill theory only for small values of the main quantum number $\bigtriangleup_0$. For example, for the permutation group $S_4$, with the replacements $(\phi_1, \phi_2)\to Z$ , $(\phi_3, \phi_5)\to W$ , $\phi_4 \to Y$ , we find
\begin{eqnarray}
z_1&=& \Tr { (2ZWZW-ZW^2Z-WZ^2W+Z^2W^2-WZWZ)Y} \quad , \nonumber\\
z_2 &=& \Tr { (W^2Z^2-2Z^2W^2+WZ^2W+ZW^2Z-WZWZ)Y} \quad , \nonumber\\
z_3 &=& \Tr { (Z^2W^2-2W^2Z^2+WZ^2W+ZW^2Z-ZWZW)Y} \quad , \nonumber\\
z_4&=& \Tr { (2WZWZ-ZW^2Z-WZ^2W+W^2Z^2-ZWZW)Y} \quad 
\quad  \nonumber
\end{eqnarray} 
One obtains one positive parity eigenstate and two negative parity eigenstates :
\begin{eqnarray}
&&\bigtriangleup_0=5 \quad , \quad \bigtriangleup_2=6 \quad , \quad {\rm parity} =+1 \nonumber\\
&& {}\qquad u=\Tr { (ZWZW+WZWZ)Y-2(ZW^2Z+WZ^2W)Y+(W^2Z^2+Z^2W^2)Y} \quad . \nonumber\\
&&\bigtriangleup_0=5 \quad , \quad \bigtriangleup_2=8 \sin^2 \frac{2\pi}{5} \quad , \quad {\rm parity} =-1 \nonumber\\
&& {}\qquad u=(z_1-z_4)\sin \frac{2\pi}{5}+(z_2-z_3)\sin \frac{4\pi}{5} \quad . \nonumber\\
&&\bigtriangleup_0=5 \quad , \quad \bigtriangleup_2=8 \sin^2 \frac{\pi}{5} \quad , \quad {\rm parity} =-1 \nonumber\\
&& {}\qquad u=(z_1-z_4)\sin \frac{4\pi}{5}+(z_3-z_2)\sin \frac{2\pi}{5} \quad . \nonumber
\end{eqnarray}

However we find that replacing  the set matrix fields $\phi_j$ with an alphabet of merely $3$ letters $Z,W,Y$ leads to vanishing basis vectors $z_j$ if the main quantum number $\bigtriangleup_0>5.$\\

\section*{Appendix D. The replacements}

Our final step is to replace the fields $\phi_j$ in the general solution ($C.1$) and ($C.2$) pertinent to the representation $(n-1,1)$ with the complex fields $Z$ , $W$ and $Y$ then obtaining the eigenvectors of the dilatation operators.\\
For the group $S_n$ such eigenvalues are sums of a large number of terms, each of them being a product of $n+1$ fields $\phi_j$. By replacing $n_1$ fields with the complex field $Z$ , $n_2$ fields with $W$ , $n_3$ fields with $Y$ , with $n_1+n_2+n_3=n+1$, one obtains sets of eigenvalues for the sector $\Tr {Z^{n_1}W^{n_2}Y^{n_3}}$.\\

The different choices of fields in the sets $n_j$ may lead  to  inequivalent sets of eigenvectors for the same sector $\Tr {Z^{n_1}W^{n_2}Y^{n_3}}$ as we indicated in Sect.4.\\
We   turn here  to the generic term in the sequences  with the simplest examples : \\
(a) we obtain the well known sequences of eigenvectors with two impurities, \cite{be}, we may call it the sector $\Tr {Z^{n-1} \phi_a \phi_b}$\\
(b) we obtain sequences of eigenvectors for the sectors $\Tr {Z^{n-2} W^2 Y}$.\\

(a). {\bf The sector $\Tr{ Z^{n-1} \phi_a \phi_b}$.}\\

By the replacement of $n-1$ matrix fields $\phi_j$ with the single complex matrix field $Z$, one recovers the well known sequences of eigenvectors with two impurities. More specifically, we identify $\{\phi_2 , \phi_3, \cdots \phi_n \}\to Z$, then
 \begin{eqnarray}
v_1& \to & (n-1)! \, \Tr {\phi_{\underline{1}}\phi_{\underline{n+1}}Z^{n-1}}
-(n-2)! \, \Tr {
\phi_{\underline{1}}Z\phi_{\underline{2}}Z^{n-2}+ \phi_{\underline{1}}Z^2\phi_{\underline{2}}Z^{n-3}+
\cdots + \phi_{\underline{1}}Z^{n-1}\phi_{\underline{2}}}= \nonumber\\
&&= 
(n-2)!\, n\, \Tr {\phi_1 \phi_{n+1}Z^{n-1}} -(n-2)!\,C \quad , \quad \nonumber\\
v_2& \to & (n-2)!\, n\, \Tr {\phi_1 Z\phi_{n+1}Z^{n-2}} -(n-2)!\,C \quad , \nonumber\\
\cdots &\to & \cdots \qquad \cdots \nonumber\\
v_n & \to & (n-2)!\, n\, \Tr {\phi_1 Z^{n-1}\phi_{n+1}} -(n-2)!\,C 
\quad  \nonumber
\end{eqnarray} 
 where $
C=\sum_{j=0}^{n-1}  \Tr {\phi_1 Z^j \phi_{n+1}Z^{n-1-j}}$. \\ 
 Since $\sum_{j=1}^n \sin \frac{2k\pi j}{n+1}=0$ and $\sum_{j=1}^n \cos \frac{k\pi (2j-1)}{n}=0$ , we recover the well known symmetric and antisymmetric sequences with two impurities
 \begin{eqnarray}
&&\bigtriangleup_2=8 \sin^2 \frac{k \pi}{n+1} \quad , \quad
u^{(k)}=\sum_{j=1}^n \sin \frac{2k\pi j}{n+1}\,\Tr {\phi_1Z^{j-1}\phi_{n+1}Z^{n-j}} \quad \nonumber\\
(D.1)&&k=1,2,\cdots ,k_{max} \quad , \quad k_{max}<\frac{n+1}{2}
\quad ,\quad {\rm parity}=-1 \quad
\nonumber
\end{eqnarray} 
and
\begin{eqnarray}
&&\bigtriangleup_2=8 \sin^2 \frac{k \pi}{n} \quad , \quad
u^{(k)}=\sum_{j=1}^n \cos \frac{k\pi (2j-1)}{n}\,\Tr {\phi_1Z^{j-1}\phi_{n+1}Z^{n-j}} \quad \nonumber\\
(D.2)&&k=1,2,\cdots ,k_{max} \quad , \quad k_{max}<\frac{n}{2}
\quad  ,\quad {\rm parity}=1 \quad
\nonumber
\end{eqnarray} 

(b). {\bf The sector $\Tr{ Z^{n-2} W^2 Y}$.}\\

Let us now look for a new sequence of eigenstates by replacing $n-2$ matrix fields $\phi_j$ with the single complex matrix field $Z$, $\{\phi_3 , \phi_4, \cdots \phi_n \}\to Z$, 
two fields with $W$, $\{\phi_1 \, ,\, \phi_2\} \to W$ , and the last one with $Y$, $\phi_{n+1}\to Y$. \\
With this replacement all the basis vector would vanish if the further replacement $Y \to W$ were made. Indeed an alternative and inequivalent replacement leading to the same sector $\Tr{ Z^{n-1} W^2 Y}$ is $\phi_1 \to Y$ , $\{\phi_2\,,\,\phi_3\} \to W$ , $\{ \phi_4,\cdots , \phi_{n+1}\} \to Z$. We limit ourselves to the first, simpler, replacement.\\
 
We find 
\begin{eqnarray}
v_1 &=&  \sum_{r=0}^{n-2} \Tr {Z^r WZ^{n-2-r}WY}-\sum_{r=0}^{n-2}  \Tr { Z^{n-2-r} W^2Z^r Y } \quad , \nonumber\\
v_2 &=& \sum_{r=0}^0 \Tr { Z^{n-2}WZ^r WZ^{-r}Y-Z^r WZ^{n-2}WZ^{-r}Y}+ \qquad  \nonumber\\
&& \qquad + \sum_{r=0}^{n-3} \Tr {Z^r WZ^{n-3-r}WZ Y-Z^r W Z W Z^{n-3-r}Y}  \quad , \nonumber\\
v_j &=&  \sum_{r=0}^{j-2} \Tr {Z^{n-j}WZ^r WZ^{j-2-r}Y-Z^r WZ^{n-j}WZ^{j-2-r}Y} + \qquad  \nonumber\\
&& \qquad +\sum_{r=0}^{n-j-1} \Tr { Z^r WZ^{n-j-1-r}WZ^{j-1}Y-Z^r WZ^{j-1}WZ^{n-j-1-r}Y}  \quad , \nonumber\\
v_{n-1} &=&  \sum_{r=0}^{n-3}  \Tr {Z W Z^r WZ^{n-3-r} Y-
Z^r W Z W Z^{n-3-r} Y} + \qquad  \nonumber\\
&& \qquad +\sum_{r=0}^{0} \Tr { Z^r W Z^{-r}WZ^{n-2}Y-
Z^r W Z^{n-2} W Z^{-r}Y} \quad , \nonumber\\
v_n &=&  \sum_{r=0}^{n-2} \Tr { W Z^r WZ^{n-2-r} Y}-\sum_{r=0}^{n-2}  \Tr { Z^{n-2-r} W^2 Z^rY } \quad 
 \quad  \nonumber
\end{eqnarray} 
 
The positive parity sequences of eigenstates , if $n$ is even , may be written
\begin{eqnarray}
&& \bigtriangleup_2=8 \sin^2 \frac{k\pi}{n} \quad , \quad
u^{(k)}=(v_1+v_n)\cos (k\pi/n)+\sum_{j=2}^{[n/2]} \cos \left( \frac{k \pi (2j-1)}{n} \right) ( v_j+v_{n-j+1})  \nonumber\\
&& (D.3) \qquad \qquad {\rm where} \qquad  v_j+v_{n-j+1} =\nonumber\\
&&\sum_{r=0}^{j-2} {\rm tr}\,[(Z^{n-j}WZ^r W Z^{j-2-r}+Z^{j-2-r}W Z^r W Z^{n-j})Y-2\,Z^r W Z^{n-j}W Z^{j-2-r}Y] + \qquad  \nonumber\\
&& +\sum_{r=0}^{n-j-1}{\rm tr}\,[ (Z^r W Z^{n-j-1-r} W Z^{j-1}+
Z^{j-1}W Z^{n-j-1-r} W Z^r)Y-2\,Z^r W Z^{j-1}W Z^{n-j-1-r}Y]  
 \qquad  \nonumber
\end{eqnarray} 
 where $k=1,2, \cdots , k_{max}$ , $k_{max}<n/2$.\\ If $n$ is odd integer, the above $u^{(k)}$ should be replaced by
$u^{(k)}+\cos (k\pi)\, v_{(n+1)/2}$ .\\

The negative parity sequences of eigenstates may be written
\begin{eqnarray}
&& \bigtriangleup_2=8 \sin^2 \frac{k\pi}{n+1} \quad , \quad
u^{(k)}=(v_1-v_n)\sin (2k\pi/(n+1))+\sum_{j=2}^{[n/2]} \sin \left( \frac{2k \pi j}{n+1} \right) ( v_j-v_{n-j+1})  \nonumber\\
&& (D.4) \qquad \qquad {\rm where} \qquad  v_j-v_{n-j+1} =\nonumber\\
&&\sum_{r=0}^{j-2} {\rm tr}\,[(Z^{n-j}W Z^r W Z^{j-2-r}-Z^{j-2-r}W Z^r W Z^{n-j})Y] + \qquad  \nonumber\\
&& +\sum_{r=0}^{n-j-1}{\rm tr}\,[ (Z^r W Z^{n-j-1-r} W Z^{j-1}-
Z^{j-1}W Z^{n-j-1-r} W Z^r)Y]  
 \quad  \nonumber
\end{eqnarray} 
 where $k=1,2, \cdots , k_{max}$ , $k_{max}<(n+1)/2$.\\

\end{document}